\documentclass[aps,prd,10pt,twocolumn,showpacs,amsmath,amssymb,nofootinbib,superscriptaddress]{revtex4-2}
\usepackage{graphicx}  
\usepackage{color}
\usepackage{times}
\usepackage{float}
\usepackage{afterpage}
\usepackage[utf8]{inputenc}
\usepackage{bm}
\usepackage{ulem}
\usepackage{multirow}
\usepackage{makecell}
\usepackage{url}
\usepackage{natbib}
\usepackage{mathrsfs}
\usepackage{physics}
\usepackage{comment}
\usepackage[table,xcdraw]{xcolor}
\usepackage{booktabs,makecell}
\usepackage{amsmath}
\usepackage{pifont}
\usepackage[colorlinks=true,citecolor=blue,urlcolor=blue,linkcolor=blue]{hyperref}
\usepackage{microtype}
\usepackage[none]{hyphenat}

\begin{document}

\title{Stability analysis of two-fluid neutron stars featuring twin star and ultradense configurations}

\author{Ankit Kumar}
\email{ankitlatiyan25@gmail.com}
\affiliation{Department of Mathematics and Physics, Kochi University, Kochi, 780-8520, Japan}

\author{Hajime Sotani}
\email{sotani@yukawa.kyoto-u.ac.jp}
\affiliation{Department of Mathematics and Physics, Kochi University, Kochi, 780-8520, Japan}
\affiliation{RIKEN Center for Interdisciplinary Theoretical and Mathematical Sciences (iTHEMS), RIKEN, Wako 351-0198, Japan}
\affiliation{Theoretical Astrophysics, IAAT, University of T\"{u}bingen, 72076 T\"{u}bingen, Germany}

\date{\today}
\begin{abstract}
We perform a detailed analysis of radial oscillations to discuss dynamical stability in two-fluid neutron stars composed of ordinary nuclear matter and a gravitationally coupled dark matter component. Using a fully relativistic two-fluid formalism, we solve the eigenvalue problem for a coupled system of equations with small-amplitude radial perturbations and derive the critical line corresponding to stability boundaries. We also compare these stability boundary lines obtained from the radial perturbations with those obtained from a generalized turning-point criterion based on extremization of mass and particle numbers, and find that the two methods agree to within better than 1\% across the parameter space explored. We consider both mirror dark matter and self-interacting fermionic dark matter models, and examine how microphysical properties---such as nuclear equations of state, dark matter mass, and vector coupling strength---reshape the topology of the stability boundary and gravitational mass contours. Our results reveal the emergence of ultra-dense and compact stars, with nuclear central densities exceeding single-fluid instability thresholds by factors of two or more, and the appearance of twin-star configurations with identical masses but distinct radii and internal fluid compositions. These findings have direct implications for the interpretation of neutron star observables and motivate future studies involving phase transitions, density discontinuities, or additional interactions in multi-component stellar systems. In particular, the emergence of exotic stable configurations beyond conventional stability limits underscores the need to reassess standard criteria in light of multi-fluid dynamics, with significant consequences for multimessenger probes of dense matter—including gravitational wave signals, mass–radius constraints, and post-merger remnants.
\end{abstract}


\maketitle

\section{Introduction}
\label{sec:1}
Neutron stars provide a unique testing ground for physics under extreme conditions, with central densities exceeding several times the nuclear saturation density. While the outer layers are composed of well-understood nuclear matter, the exact nature and behavior of the high-density core remain uncertain due to the poorly constrained equation of state (EOS) at supranuclear densities~\cite{Bogdanov_2019, Bogdanov_2021}. In recent years, the possibility that dark matter (DM)---a non-luminous component that constitutes most of the matter in the universe~\cite{Jarosik_2011, ZUREK201491}---may coexist with nuclear matter inside neutron stars has attracted significant attention~\cite{PhysRevD.92.063526, PhysRevD.104.063028, galaxies10010014, Sagun_2023, PhysRevD.111.043016, particles7010011, PhysRevD.111.083038}. If DM accumulates during stellar evolution or is present when a star forms, it can alter the star's structure, stability, and oscillation properties, potentially affecting observable quantities, including thermal emission properties and gravitational wave signatures~\cite{PhysRevD.111.043016, 10.1093/mnras/stae337, PhysRevD.103.123022, 10.1093/mnras/stac1013, 10.1093/mnras/stab2387, 10.1093/mnrasl/slv049}. A widely adopted theoretical framework for studying such composite stars is the two-fluid model, where nuclear and DM components are treated as distinct fluids coupled only through gravity. Depending on the relative central densities and internal properties of the DM fluid, DM can form a compact core, an extended halo, or a distribution comparable to the nuclear component~\cite{Giangrandi_2023, Nelson_2019, kumar2025mcc, PhysRevD.111.043050}. These configurations can have measurable effects on astrophysical observables, and also fundamentally impact the dynamical stability of the star ~\cite{PhysRevD.40.3221}.

In neutron stars, radial oscillations---small perturbations preserving spherical symmetry---provide a fundamental tool to probe stellar stability and internal structure. For single-fluid neutron stars, the analysis of radial oscillations is well-established: small radial perturbations in pressure, density, and the metric about an equilibrium configuration lead to a second-order Sturm-Liouville eigenvalue problem~\cite{1964ApJ...140..417C, 1977ApJ...217..799C, KokkotasRuoff, PhysRevLett.77.4134}. The solutions yield discrete eigenfrequencies associated with distinct radial modes. A neutron star configuration is dynamically stable if all squared-eigenfrequencies are positive; conversely, the onset of dynamical instability corresponds to the point at which the fundamental mode frequency vanishes~\cite{10.1111/j.1365-2966.2006.11304.x}. This criterion coincides precisely with the turning point in the mass-central energy density ($M$-$\varepsilon_c$) curve, where the condition $dM/d\varepsilon_c = 0$ marks the threshold between stable and unstable configurations~\cite{1988ApJ...325..722F, 1996JApA...17..199F, l9p9-nqbx}. Beyond this turning point, radial perturbations cannot restore equilibrium, and the star becomes dynamically unstable, potentially leading to gravitational collapse.

However, extending radial oscillation theory and associated stability criteria from single-fluid to two-fluid neutron star models is non-trivial and remains relatively unexplored. In a general two-fluid formalism, each fluid component follows its own EOS, and the dynamics depend significantly on how these fluids interact. Broadly, two classes of two-fluid systems have been considered in literature: (i) fluids interacting through additional non-gravitational couplings, such as entrainment effects in superfluid neutron stars, described typically by a master function formalism~\cite{1988ApJ...333..880E, PhysRevD.60.104025, 10.1046/j.1365-8711.2001.04923.x, PhysRevD.66.104002, 10.1111/j.1365-2966.2008.13426.x}; and (ii) fluids interacting solely through gravitational coupling~\cite{GOLDMAN2013200, SANDIN2009278, PhysRevD.105.123010, PhysRevD.111.083038, rutherford2024, kumar2025mcc}. The presence or absence of non-gravitational interactions fundamentally alters the equations governing the equilibrium structure, perturbation dynamics, and associated stability conditions. In the present study, we focus exclusively on the latter scenario, considering two fluids coupled only gravitationally. Even in this gravitationally coupled case, stability conditions depend not only on the total stellar mass and central densities but also explicitly on the compositions and relative distributions of the individual fluid components. 
The existence of an additional fluid introduces new degrees of freedom, resulting in a coupled system of perturbation equations that must be solved simultaneously to determine the radial mode frequencies and corresponding eigenfunctions. Therefore, developing a rigorous two-fluid stability analysis necessitates careful derivation of these perturbation equations; the implementation of appropriate boundary conditions at the center, the interface of the fluid regions, and the stellar surface; and a systematic examination of how gravitational interactions between the fluids influence the onset of dynamical stability.

While solving the coupled radial pulsation equations provides a rigorous means to determine the dynamical stability of two-fluid stars, an alternative and computationally efficient method is offered by a generalized turning-point criterion. For single-fluid stars, the transition from stable to unstable configurations is uniquely determined by the maximum mass point along a sequence parametrized by the central density. However, in the two-fluid case---where equilibrium configurations form a two-parameter family, characterized, for example, by the central densities (or pressures) of each fluid---there is no unique path through parameter space, and the notion of a turning point becomes inherently multidimensional in parameter space. This geometric method (i.e. generalized turning-point method) was originally formulated by Henriques, Liddle, and Moorhouse in the context of fermion-boson stars~\cite{HENRIQUES1990511}, where they demonstrated that the stability boundary corresponds to configurations where both the total mass and the individual particle numbers of each fluid simultaneously extremize. It has been subsequently applied to DM admixed neutron stars~\cite{PhysRevD.103.043009, PhysRevD.111.043050}, providing an effective tool to identify the onset of instability based solely on equilibrium sequences, without requiring the solution of the coupled perturbation equations. In the present work, we implement both approaches---the eigenmode analysis with radial perturbations and the generalized turning-point method---to map the stability boundaries in two-fluid neutron star models, providing a systematic comparison between these complementary criteria.

Understanding the stability boundaries of two-fluid neutron stars is not only of theoretical interest but also has direct implications for astrophysical observations. Shifts in the stability boundary can modify the maximum mass, compactness, and tidal deformability allowed for a given EOS, thereby influencing the interpretation of pulsar timing, X-ray measurements, and gravitational wave signals from neutron star mergers. Although radial oscillation modes do not emit gravitational waves due to their preservation of spherical symmetry, and direct electromagnetic observability remains elusive, nonlinear coupling between radial and nonradial modes may produce complex surface phenomena that could, in principle, manifest as observable electromagnetic variability~\cite{PhysRevD.65.024001, PhysRevD.73.084010}. Consequently, robust stability analysis provides a crucial link between microscopic particle physics models and the macroscopic observables accessible to current and future multi-messenger astronomy. A detailed understanding of two-fluid stability is thus essential for interpreting DM scenarios in compact stars.

The paper is structured as follows: In Sec.~\ref{sec:2}, we present the theoretical framework, beginning with the formulation of radial oscillations in two-fluid neutron stars (Sec.~\ref{sec:2A}), followed by the generalized turning-point method for assessing dynamical stability (Sec.~\ref{sec:2B}), and concluding with a description of the EOS used for the nuclear and DM components (Sec.~\ref{sec:2C}). Section~\ref{sec:3} presents a detailed analysis of our numerical results. We analyze the frequency structure and stability boundaries across a range of DM models, i.e., the mirror DM (Sec.~\ref{sec:3A}) and the self-interacting fermionic DM models (Sec.~\ref{sec:3B}). Then, we explore the emergence of ultra-dense configurations and twin-star structures in specific regions of the parameter space in Sec.~\ref{sec:4}. Finally, in Sec.~\ref{sec:5}, we summarize our key findings and discuss their broader implications in the context of neutron star astrophysics and DM phenomenology.

\section{Formalism}
\label{sec:2}
\subsection{Radial oscillations in two-fluid stars}
\label{sec:2A}
We consider a static, spherically symmetric relativistic star composed of two distinct fluids, here denoted as fluid X and fluid Y, which interact solely through gravity. This setup is applicable to a broad class of models, including (but not limited to) DM admixed neutron stars, as long as additional non-gravitational interactions such as entrainment or chemical couplings can be neglected. In particular, we assume that the two fluids are dynamically decoupled, meaning their individual stress-energy tensors, $T_{(i)}^{\mu\nu}$, are separately conserved, i.e., $\nabla_{\nu} T^{\mu\nu}_{(i)} = 0$ for each $i = X, Y$. This assumption excludes entrainment effects, where the momentum of one fluid depends on the motion of the other.

The background spacetime geometry is described by the standard Schwarzschild-like metric,
\begin{equation}
ds^2 = -e^{2\Phi(r)}dt^2 + e^{2\Lambda(r)}dr^2 + r^2\ d\theta^2 + r^2\ \sin^2\theta\ d\phi^2, \nonumber
\end{equation}
where $\Phi(r)$ and $\Lambda(r)$ are the metric potentials determined by the matter distribution within the star. In particular, \( \Phi(r) \) represents the gravitational potential affecting redshift and time dilation, while \( \Lambda(r) \) is related to the curvature of the radial part of the geometry and is often expressed in terms of the enclosed (total) mass function. Each fluid is modeled as a perfect fluid, described by its energy density ${\cal E}_{i}$, pressure $P_{i}$, and four-velocity $u^{\mu}_{i}$ ($i = X, Y$), and contributes additively to the total stress-energy tensor
\begin{equation}
T^{\mu\nu} = T^{\mu\nu}_{(X)} + T^{\mu\nu}_{(Y)}
\end{equation}
with
\begin{equation}
T^{\mu\nu}_{(i)} = ({\cal E}_i + P_i)\ u_i^\mu u_i^\nu + P_{i}\ g^{\mu\nu}\ .
\end{equation}
To construct the equilibrium structure of a two-fluid star, we solve Einstein's field equations in conjunction with the energy-momentum conservation laws imposed separately on each fluid component. Under the assumptions of staticity and spherical symmetry, the conservation equations reduce to a single non-trivial condition per fluid---the radial component---which yields the relativistic hydrostatic equilibrium equation:
\begin{equation}
    \frac{dP_i}{dr} = -\left(\mathcal{E}_i + P_i\right) \frac{d\Phi}{dr}\ .
    \label{eq:eq3}
\end{equation}
This relation governs how each fluid maintains its pressure gradient to counterbalance the inward gravitational pull encoded in $\Phi(r)$. The spacetime curvature in the radial direction is encoded by the metric function $\Lambda(r)$, which relates to the cumulative gravitational mass enclosed within radius $r$ through $e^{-2\Lambda(r)} = 1 - 2\ m(r) / r$. Here, $m(r)$ is the total mass function, obtained by integrating the combined energy density contributions of both fluids:
\begin{equation}
    \frac{dm}{dr} = 4\pi r^{2} \sum_{i=X,Y} \mathcal{E}_i\ .
    \label{eq:eq4}
\end{equation}
Finally, the radial gradient of the gravitational potential follows from the $rr-$component of Einstein's equations and is given by
\begin{equation}
    \frac{d\Phi}{dr} = \frac{1}{r(r - 2m)} \left(m + 4\pi r^3 \sum_{i=X,Y} P_i\right),
\label{eq:eq5}
\end{equation}
The equilibrium background is governed by the two-fluid Tolman-Oppenheimer-Volkoff (TOV) equations [i.e., Eqs.~\eqref{eq:eq3}-\eqref{eq:eq5}], where each fluid separately satisfies its own hydrostatic equilibrium condition in the shared gravitational potential~\cite{kcl2-qgxh}. These equations constitute a system for six variables---$\mathcal{E}_X$, $P_X$, $\mathcal{E}_Y$, $P_Y$, $\Phi$, and $m$---and require two independent EOS, one for each fluid, to close the system. The specific forms of the EOS for the nuclear and DM components are described in Sec.~\ref{sec:2C}.

Now, to study small-amplitude radial oscillations, we introduce independent Lagrangian displacements, $\xi_{X}(r, t)$ and $\xi_{Y}(r, t)$, which describe the radial motion of fluid elements about equilibrium. The linearized dynamics are obtained by expanding both the conservation equations for each fluid ($\nabla_{\nu} T^{\mu\nu}_{(i)} = 0$) and the Einstein field equations to first order in the perturbations. In this formalism, all perturbed fluid variables and metric potentials are expressed in terms of the displacement fields, and the gravitational interaction between fluids manifests naturally via metric perturbations sourced by their combined stress-energy tensor, reflecting the shared spacetime geometry.

The derivation of the coupled radial pulsation equations for a two-fluid star is a non-trivial task, requiring careful treatment of both the perturbed fluid and the Einstein field equations, as well as the interplay between Eulerian and Lagrangian perturbations. For the detailed derivation, we refer the reader to Refs.~\cite{PhysRevD.110.103038, PhysRevD.102.023001}; in particular, we closely follow the methodology and conventions in Ref.~\cite{PhysRevD.110.103038}. The final outcome is a set of two coupled second-order differential equations for the displacement fields, $\xi_X(r, t)$ and $\xi_Y(r, t)$, capturing both the individual dynamics of each fluid and their mutual coupling through the shared spacetime geometry. For completeness and clarity, we present below the explicit form of the radial pulsation equation used in our calculations, noting that while its general structure is guided by the cited literature, we have derived and arranged it in a form specifically suited to our analysis of two-fluid stars. Equation governing the time evolution of the Lagrangian displacement, $\xi_X(r, t)$, for fluid $X$ is
\begin{widetext}
\begin{align}
    e^{2(\Lambda-\Phi)} & \left({\cal E}_{X}+P_{X}\right) \frac{\partial^{2} \xi_{X}}{\partial t^{2}} = 
    \notag \\
    &-\frac{\partial}{\partial r} \left[\xi_{X} \left({\cal E}_{X}+P_{X} \right)\frac{d\Phi}{dr} - \gamma_{X} P_{X}\frac{e^{\Phi}}{r^{2}}\frac{\partial}{\partial r}\left(r^{2}\xi_{X} e^{-\Phi}\right) + 4\pi r e^{2\Lambda}\gamma_{X}P_{X} \left({\cal E}_{Y}+P_{Y}\right) \left(\xi_{Y}-\xi_{X}\right)\right] 
    \notag \\
    & - \left(1+ \frac{{\cal E}_{X}+P_{X}}{\gamma_{X} P_{X}}\right) \frac{d\Phi}{dr} \left[\xi_{X} \left({\cal E}_{X}+P_{X}\right) \frac{d\Phi}{dr} - \gamma_{X} P_{X} \frac{e^{\Phi}}{r^{2}}\frac{\partial}{\partial r}\left(r^{2}\xi_{X} e^{-\Phi}\right) + 4\pi r e^{2\Lambda}\gamma_{X}P_{X} \left({\cal E}_{Y}+P_{Y}\right) \left(\xi_{Y}-\xi_{X}\right)\right] \notag \\
    & + 4\pi re^{2\Lambda} \left({\cal E}_{X}+P_{X}\right) \left(\frac{d\Phi}{dr}+\frac{1}{r}\right) \Big[\xi_{X} \left({\cal E}_{X}+P_{X}\right) + \xi_{Y} \left({\cal E}_{Y}+P_{Y}\right)\Big] \notag \\
    & + 4\pi re^{2\Lambda} \left({\cal E}_{X}+P_{X}\right) \frac{e^{\Phi}}{r^{2}} \left[\gamma_{X} P_{X} \frac{\partial}{\partial r} \left(r^{2}\xi_{X} e^{-\Phi}\right) + \gamma_{Y} P_{Y} \frac{\partial}{\partial r} \left(r^{2}\xi_{Y} e^{-\Phi}\right)\right]
    \notag \\
    & - \left(4\pi re^{2\Lambda}\right)^{2} \left({\cal E}_{X}+P_{X}\right) \Big[\gamma_{X}P_{X} \left({\cal E}_{Y}+P_{Y}\right) \left(\xi_{Y}-\xi_{X}\right) + \gamma_{Y}P_{Y} \left({\cal E}_{X}+P_{X}\right) \left(\xi_{X}-\xi_{Y}\right) \Big], \label{eq:rad_pert_eq}
\end{align}
\end{widetext}
which is coupled with the Lagrangian displacement, $\xi_Y(r,t)$, for the fluid $Y$.
Here, the adiabatic index $\gamma_X$ is defined as $\gamma_X = ({\cal E}_X + P_X)/P_X \cdot (\partial P_X/\partial {\cal E}_X)$, and all background quantities---such as ${\cal E}_X$, ${\cal E}_Y$, $P_X$, $P_Y$, $\Phi(r)$, and $\Lambda(r)$---are evaluated in the equilibrium configuration. The above equation together with the corresponding equation for $\xi_Y(r, t)$ (obtained by exchanging $X \leftrightarrow Y$), describes the coupled dynamics of radial perturbations in both fluids. 

To facilitate numerical integration, it is convenient to recast these coupled perturbation equations as an eigenvalue problem for discrete radial oscillation modes. To this end, we first adopt the standard decomposition of the perturbations into temporal and spatial components as
\begin{equation}
\xi_{X}(r,t) = \eta_{X}(r)\ \Theta_{X}(t), \quad \xi_{Y}(r,t) = \eta_{Y}(r)\ \Theta_{Y}(t).
\end{equation}
Due to the gravitational coupling and the linear nature of the perturbation equations, both fluids must exhibit the same temporal behavior, apart from a possible constant phase factor. Thus, we can write $\Theta_{X}(t) = \Theta_{Y}(t) \propto e^{i\omega t}$, where $\omega$ denotes the common angular eigenfrequency of oscillation for both fluids. With this temporal decomposition, the radial displacement fields now take the simpler harmonic form:
\begin{equation}
\xi_{X}(r,t) = e^{i\omega t}\ \eta_{X}(r), \quad \xi_{Y}(r,t) = e^{i\omega t}\ \eta_{Y}(r).
\end{equation}
To further simplifying the equations and enhance numerical convenience, we introduce new radial displacement variables $\zeta_{X}(r)$ and $\zeta_{Y}(r)$, defined by
\begin{equation}
\zeta_{X} = r^{2}e^{-\Phi}\eta_{X}, \quad \zeta_{Y} = r^{2}e^{-\Phi}\eta_{Y}.
\end{equation}
Applying these transformations and definitions, the coupled radial perturbation equations can be reduced-after detailed algebraic simplifications (see the Appendix of ref.~\cite{PhysRevD.110.103038} for details):
\begin{align}
    \omega^{2}A_{X}\ \zeta_{X} =& -\frac{d}{dr} \left(W_{X}\frac{d\zeta_{X}}{dr}\right) \ +\  \big[ Q_{X}+T \big]\,\zeta_{X} +\ S\,\zeta_{Y} \nonumber \\
    & + \big(\zeta_{Y} - \zeta_{X}\big) \frac{d f_{X}}{dr} + \big(f_{X} - f_{Y}\big) \frac{d \zeta_{Y}}{dr},
    \label{eq:eq10}
\end{align}
alongside a similar coupled equation for fluid $Y$, obtained by interchanging subscripts $X \leftrightarrow Y$. Here, the coefficients $A_{X},\ W_{X},\ Q_{X},\ T,\ S,\ f_{X}$ and $f_{Y}$ are expressed explicitly in terms of background equilibrium quantities as follows:
\begin{align}
    A_{X} = &\ e^{3\Lambda+\Phi}\frac{({\cal E}_{X}+P_{X})}{r^{2}}, \nonumber \\
    W_{X} = &\ e^{\Lambda+3\Phi} \frac{\gamma_{X} P_{X}}{r^{2}}, \nonumber \\
    Q_{X} = &\ e^{\Lambda+3\Phi}\frac{{\cal E}_{X}+P_{X}}{r^{2}} \left[\frac{d^{2}\Phi}{dr^{2}}-\frac{2}{r}\frac{d\Phi}{dr} + \left(\frac{d\Phi}{dr}\right)^{2}\right] \nonumber \\
    &\ -4\pi e^{3\Lambda+3\Phi} \frac{\left({\cal E}_{X}+P_{X}\right)^{2}}{r} \left(\frac{d\Phi}{dr}+\frac{1}{r}\right), \nonumber \\
    T = &\ -4\pi \frac{e^{3\Lambda+3\Phi}}{r}\frac{d\Phi}{dr} \left({\cal E}_{X}+P_{X}\right) \left({\cal E}_{Y}+P_{Y}\right) \nonumber \\
    &\ + 16\pi^{2} e^{5\Lambda+3\Phi}\gamma_{X}P_{X} \left({\cal E}_{Y}+P_{Y}\right)^{2} \nonumber \\
    &\ + 16\pi^{2} e^{5\Lambda+3\Phi}\gamma_{Y}P_{Y} \left({\cal E}_{X}+P_{X}\right)^{2}, \nonumber \\
    S = &\ -4\pi \frac{e^{3\Lambda+3\Phi}}{r^{2}} \left({\cal E}_{X}+P_{X}\right) \left({\cal E}_{Y}+P_{Y}\right) \nonumber \\ 
    &\ - 16\pi^{2} e^{5\Lambda+3\Phi}\gamma_{X}P_{X} \left({\cal E}_{Y}+P_{Y}\right)^{2} \nonumber \\
    &\ - 16\pi^{2} e^{5\Lambda+3\Phi}\gamma_{Y}P_{Y} \left({\cal E}_{X}+P_{X}\right)^{2}, \nonumber \\
    f_{X} = &\ \frac{4\pi}{r} e^{3\Lambda+3\Phi}\gamma_{X}P_{X} \left({\cal E}_{Y}+P_{Y}\right), \nonumber \\
    f_{Y} = &\ \frac{4\pi}{r} e^{3\Lambda+3\Phi}\gamma_{Y}P_{Y} \left({\cal E}_{X}+P_{X}\right). \nonumber
\end{align}
Equation~\eqref{eq:eq10} and its fluid-$Y$ counterpart constitute the complete set of coupled equations governing the radial oscillations of two-fluid neutron stars. 

In order to solve these eigenvalue equations, one must impose boundary conditions at three locations: the stellar center, the inner fluid surface (inner boundary), and the outermost stellar surface. These conditions arise naturally from the requirements of regularity at the center and continuity of fluid and metric variables across the inner boundary and at the surface.
\begin{itemize}
    \item At the center of the star ($r=0$), physical regularity requires that the radial displacement fields do not diverge. This leads to the conditions:
    \begin{align}
    \lim_{r \rightarrow 0} \frac{\xi_{X}(r)}{r} \approx \text{finite}, \quad
    \lim_{r \rightarrow 0} \frac{\zeta_{X}(r)}{r^{3}} = k_X\ ,
    \end{align}
    with some constant, $k_X$, ensuring a smooth and non-singular behavior near the star's center. A similar condition applies to fluid $Y$ as well, characterized with $k_Y$. Owing to the nature of linear analysis, one can choose $k_X=1$ as the normalized condition.
    \item At the inner boundary, where the inner fluid $X$ vanishes (i.e., at $r = R_X$)\footnote{In the present analysis, we explicitly assume that fluid $X$ represents the inner fluid, occupying the region from the stellar center ($r = 0$) up to its own surface at $r = R_{X}$. The second fluid, fluid $Y$, is treated as the outer fluid: it coexists with fluid $X$ in the region $0 \le r \le R_X$, and then continues as the only constituent from $r = R_X$ to the outermost stellar surface at $r = R_Y = R$.}, the Lagrangian pressure perturbation for fluid $X$ must vanish, i.e., $\Delta P_X (R_X) = 0$. However, the presence of the non-vanishing fluid $Y$ modifies the condition, resulting in:
\begin{align}
    \gamma_{X} P_{X} \frac{e^{\Phi}}{r^{2}}\left[-\frac{d\zeta_{X}}{dr}
    + 4\pi r e^{2\Lambda}({\cal E}_{Y}+P_{Y})(\zeta_{Y}-\zeta_{X})\right] \Bigg| _{R_{X}} = 0. \nonumber
\end{align}
As $P_X \to 0$ at $R_X$, the overall Lagrangian pressure perturbation tends to zero unless there are divergences in the bracketed terms. To ensure a well-behaved solution, we require that $\zeta_X$, $d\zeta_X/dr$, and $\zeta_Y$ remain finite at the interface. Furthermore, continuity demands that both $\zeta_Y$ and its radial derivative $d\zeta_Y/dr$ are continuous across $r = R_X$.
\item At the outer boundary ($r = R_Y$), where $P_Y \to 0$, the boundary condition similarly requires that the Lagrangian pressure perturbation for fluid $Y$ vanishes:
\begin{align}
    \Delta P_{Y} = -\gamma_{Y} P_{Y} \frac{e^{\Phi}}{r^{2}} \frac{d\zeta_{Y}}{dr}\Bigg|_{r = R_{Y}} = 0. \nonumber
\end{align}
However, since $P_{Y} \rightarrow 0$ at the surface, this condition does not constrain the eigenvalue directly but implies that $d\zeta_{Y}/dr$ remains finite. Physically, the outer boundary condition thus requires that the radial displacement function $\zeta_{Y}$ approaches a constant in the vicinity of the stellar surface. This ensures the absence of diverging forces and is consistent with a smooth matching to the vacuum exterior.
\end{itemize} 

To numerically solve the eigenvalue problem defined by Eq.~\eqref{eq:eq10} and its fluid-$Y$ counterpart, we adopt a shooting method. For a given trial eigenvalue $\omega^{2}$, we integrate the coupled perturbation equations starting from the center, imposing regularity conditions on the displacement functions. In our numerical calculations, we treat the value of $\zeta_{Y}=k_Y$ at the center as a shooting parameter, adjusting it iteratively to satisfy the inner boundary condition at $r = R_{X}$. Once a solution consistent with the inner boundary condition is obtained, the corresponding values of all perturbation variables are used for further integration outward through the outer fluid layer, continuing up to the stellar surface at $r = R_{Y}$. The integration is completed at the stellar surface, where the outer boundary condition $\Delta P_{Y} (R_{Y}) = 0$ must be exactly satisfied; this serves as the final and decisive criterion for determining whether the chosen trial eigenvalue $\omega^{2}$ corresponds to a true physical oscillation mode.

For each equilibrium stellar configuration, specified by a pair of central densities (${\cal E}_{c}^{X}$, ${\cal E}_{c}^{Y}$), the requirement that all boundary conditions are simultaneously satisfied yields a discrete spectrum of allowed eigenfrequencies, $\omega^{2}$. These correspond to the natural radial oscillation modes of the two-fluid star. The fundamental mode---characterized by the lowest $\omega^{2}$, which is the mode without any nodes in the corresponding eigenfunction---serves as a principal indicator of dynamical stability. A positive $\omega^{2}$ implies that small radial perturbations result in stable, periodic oscillations about equilibrium. Conversely, a negative value of $\omega^{2}$ signals that the star is dynamically unstable to radial perturbations and may undergo exponential growth of perturbations or collapse into a black hole. While linear perturbation analysis cannot determine the precise outcome, the sign of $\omega^{2}$ provides a definitive criterion for the (in)stability of the equilibrium configuration.

\subsection{Generalized turning-point stability method}
\label{sec:2B}
A complementary approach to the eigenmode analysis for assessing dynamical stability is provided by the generalized turning-point method, originally introduced by Henriques, Liddle, and Moorhouse~\cite{HENRIQUES1990511}. This method generalizes the classical stability criterion based on mass extrema in single-fluid stars to systems composed of multiple fluids, described by multiple conserved quantities.

In single-fluid stars, stability is governed by a unique sequence of equilibrium configurations parameterized by the central energy density or pressure. Along such a sequence, the point of maximum gravitational mass marks the onset of instability. However, a similar transition also occurs at sufficiently low central densities, where stars may again become dynamically unstable---corresponding to a lower boundary in the stability domain. Both of these turning points can, in principle, be identified through extrema in the gravitational mass.

In two-fluid stars, the situation is more complex: equilibrium configurations span a two-dimensional parameter space, usually described by the central densities (${\cal E}_{c}^{X}$, ${\cal E}_{c}^{Y}$) of each fluid. In such a multidimensional parameter space, there exists no unique ordering of mass configurations, and the traditional one-dimensional turning-point criterion must be generalized. The generalized method identifies stability boundaries as surfaces where all conserved quantities---total gravitational mass $M\ (=M_{X}+M_{Y})$ with $M_X$ and $M_Y$ being the gravitational masses of fluid $X$ and $Y$, and the individual particle numbers $N_{X}$ and $N_{Y}$---are simultaneously extremized \cite{PhysRevD.87.084040}. Physically, this corresponds to configurations, where small variations in the central densities do not induce any first-order change in $M$, $N_{X}$, or $N_{Y}$, i.e.,
\begin{equation}
    \delta M = 0, \quad \delta N_{X} = 0, \quad \delta N_{Y} = 0.
\end{equation}
Mathematically, these conditions define a co-dimension one hypersurface in the (${\cal E}_{c}^{X}$, ${\cal E}_{c}^{Y}$) space, separating stable and unstable regions. The presence of a zero-frequency mode in the radial perturbation, i.e., $\omega^{2}= 0$, at such a point ensures that the configuration sits at a local extremum of these conserved quantities \cite{PhysRevD.107.115028}.

This geometric approach thus circumvents the need for solving the full set of coupled radial perturbation equations. Instead, by computing sequences of equilibrium configurations and tracking the extrema of $M$, $N_{X}$, and $N_{Y}$, one can efficiently delineate the stability boundary. In the present work, we also employ this method alongside the eigenmode analysis to systematically chart the stability boundaries of two-fluid stars---including not only the high-density (right-edge in the ${\cal E}_{c}^{X}-{\cal E}_{c}^{Y}$ plane) turning points already studied in prior works~\cite{PhysRevD.103.043009, PhysRevD.111.043050, PhysRevD.105.023010, PhysRevD.110.023013, PhysRevD.102.084063, particles7010004}, but also the low-density (left-edge) boundary, which we identify here for the first time in this context.

To implement the generalized turning-point method, one must evaluate not only the total gravitational mass but also the individual fluid particle numbers, $N_{X}$ and $N_{Y}$, across equilibrium configurations. These quantities are not simply baryon numbers, but rather represent the number of fluid elements associated with each conserved current. To compute the conserved particle numbers, $N_{X}$ and $N_{Y}$, we adopt the general relativistic definition of the total number of fluid elements in the local rest frame~\cite{PhysRevD.40.327, PhysRevD.104.043001}:
\begin{equation}
    \frac{dN_{i}}{dr} = 4\pi r^{2} n_{i}\left(1-\frac{2m}{r}\right)^{-1/2},
\end{equation}
where $n_{i}(r)$ is the proper number density of the $i$-th fluid. For isentropic fluids, the conservation law, i.e., $\left(n_{i}u^{\mu}\right)_{;\mu}$, combined with thermodynamic consistency, implies the following differential relation:
\begin{equation}
\frac{dn_i}{n_i} = \frac{d\mathcal{E}_i}{\mathcal{E}_i + P_i}\ ,
\end{equation}
which leads to
\begin{equation}
n_{i} \propto \exp\left( \int \frac{d{\cal E}_{i}}{{\cal E}_{i} + P_{i}} \right)\ .
\end{equation} 
This expression, while accurate up to a normalization constant, is sufficient for tracking relative variations in $N_{i}$ and identifying extrema along a sequence of equilibrium configurations (see Ref.~\cite{PhysRevD.40.327} for a full derivation). In our implementation, we evaluate these integrals numerically during the background TOV integration, allowing us to monitor the behavior of 
$N_{X}$ and $N_{Y}$ in the two-dimensional parameter space of central energy densities.

\subsection{Equation of state}
\label{sec:2C}
In this work, we model neutron stars as two-fluid systems composed of ordinary nuclear matter and a gravitationally coupled DM component. While several scenarios for DM microphysics have been proposed~\cite{cirelli2024darkmatter}, we focus on two representative models to investigate their impact on the equilibrium structure and dynamical stability of two-fluid stars.

As a baseline, we consider a minimal mirror DM scenario~\cite{Kolb1985, doi:10.1142/S0217751X04020075, doi:10.1142/S0217751X14300130, PhysRevLett.96.081801, PhysRevD.68.023518, PhysRevD.78.123003, SANDIN2009278, Goldman_2011}, in which the dark sector is assumed to be a thermodynamic copy of the visible sector. In this case, the DM fluid follows the same EOS as nuclear matter, but is gravitationally coupled and specified by an independent central energy density. Although this mirror DM setup does not introduce additional microphysical parameters, it serves as a useful reference framework for testing two-fluid effects in a minimal and self-consistent setting.

To describe the nuclear matter component, we employ three representative EOSs spanning a wide range of microphysical models and stiffness:
\begin{itemize}
    \item QMC-RMF4~\cite{PhysRevC.106.055804}: This EOS is formulated within the relativistic mean-field (RMF) framework whose couplings are calibrated to reproduce quantum Monte Carlo (QMC) results for pure neutron matter obtained from chiral effective field theory ($\chi$EFT). The RMF coupling parameters are tuned to match the $\chi$EFT band over the range \(0.5\,n_0\) to \(2\,n_0\) and to reproduce key properties of symmetric nuclear matter at saturation, such as the binding energy and incompressibility. These features yield a moderately stiff high-density behavior consistent with the \(2\,M_\odot\) mass constraint. In the single-fluid limit---i.e., for an isolated neutron star composed solely of nuclear matter---the QMC-RMF4 EOS predicts a maximum gravitational mass of $M = 2.206\ M_{\odot}$, with a corresponding radius of $R = 10.95$ km. This configuration is attained at a central energy density of $\approx 1290\ \mathrm{MeV/fm^3}$, marking the threshold for dynamical stability in the absence of DM.
    \item DD2~\cite{PhysRevC.81.015803, PhysRevC.89.064321}: Based on a density-dependent relativistic mean-field approach, the DD2 EOS is constrained by nuclear saturation properties and finite-nucleus data. Its relatively stiff behavior at supranuclear densities supports a maximum neutron star mass of $M = 2.427\ M_{\odot}$ with a radius of $R = 11.92$ km in the single-fluid star scenario. The corresponding central energy density at the mass peak is $\approx 1090\ \mathrm{MeV/fm^3}$.
    \item QHC21-BT~\cite{Kojo_2022}: This EOS incorporates a smooth crossover from hadronic matter, described by Togashi EOS~\cite{TOGASHI201778}, to deconfined quark matter using a Nambu-Jona-Lasinio-type model. The Togashi EOS is constructed using a variational method with realistic two-body and empirical three-body nuclear forces, and provides a unified treatment of both uniform nuclear matter and non-uniform crust matter across a wide density range. The overall QHC21 construction satisfies both nuclear saturation and QCD constraints, while ensuring causal behavior at high densities. In the single-fluid star case, the maximum mass predicted is $M = 2.200\ M_{\odot}$ with a radius of $R = 11.06$ km and central energy density around $1234\ \mathrm{MeV/fm^3}$.

\end{itemize}

While the mirror DM matter scenario is explored using all three EOSs listed above, our main focus in this work is on a physically motivated fermionic DM model featuring self-interactions mediated by a vector boson. In this setup, the dark sector consists of a Dirac fermion, $\chi$, charged under a hidden $U(1)_{V}$ gauge symmetry, with gauge coupling, $g_{\chi}$, and a corresponding vector mediator of mass, $m_{v}$. The Lagrangian governing this dark sector is given by~\cite{PhysRevD.99.083008, kumar2025mmc}:
\begin{equation}
\mathcal{L}_{\rm DM} = \bar\chi(i\gamma^\mu D_{\mu} - m_\chi)\chi - \frac{1}{2} m_{\rm v}^2 V^\mu V_\mu - \frac{1}{4} Z^{\mu\nu}Z_{\mu\nu},
\end{equation}
where $D_\mu = \partial_\mu + i g_\chi V_\mu$ is the gauge-covariant derivative and $Z_{\mu\nu}$ is the field strength of the dark vector field. 

Assuming a cold and degenerate Fermi gas of DM particles under mean-field approximation, the energy density and pressure of the vector DM fluid are given by:
\begin{align}
    {\cal E}_{\rm{DM}} &= \frac{2}{\left(2\pi\right)^{3}} \int^{k_{\chi}^{F}}_{0} \sqrt{k^{2}+m_{\chi}^{2}}\, d^{3}k \,+\,  \frac{1}{2} \left(\frac{g_{\chi}}{m_{\rm{v}}}\right)^{2} n_{\chi}^{2}, \nonumber \\
    P_{\rm{DM}} &= \frac{2}{3\left(2\pi\right)^{3}} \int^{k_{\chi}^{F}}_{0} \frac{k^{2}}{\sqrt{k^{2}+m_{\chi}^{2}}}\, d^{3}k \,+\,  \frac{1}{2} \left(\frac{g_{\chi}}{m_{\rm{v}}}\right)^{2} n_{\chi}^{2}, \label{eq:DM_EOS}
\end{align}
where $n_{\chi}$ is the number density of dark fermions and $k^{F}_{\chi}$ is the corresponding Fermi momentum. The first terms represent the kinetic and rest mass contributions with the DM particle mass, $m_\chi$, while the second terms capture the repulsive self-interaction mediated by the light vector field. One can find from Eq.~(\ref{eq:DM_EOS}) that the DM EOS becomes softer with larger $m_\chi$.

This vector-mediated self-interacting DM model provides a theoretically and phenomenologically rich alternative to the mirror matter scenario. It introduces an adjustable self-coupling strength through the ratio $g_{\chi}/m_{v}$, and is well-motivated in the context of asymmetric DM frameworks~\cite{PhysRevD.111.123034}, where the relic abundance arises from a particle–antiparticle asymmetry similar to the baryon sector. In such models, the observed abundance of the dark matter and baryons in the present Universe, i.e., $\Omega_{\rm DM}/\Omega_{b} \simeq 5.4$~\cite{ParticleDataGroup:2024cfk}, naturally points to a dark fermion mass in the few-GeV range. Accordingly, we focus on representative DM masses between 1 and 10 GeV, spanning the parameter space favored by asymmetric DM cosmologies.

The choice of the coupling ratio $g_{\chi}/m_{v}$ is guided by both theoretical considerations and astrophysical constraints. This parameter controls the stiffness of the DM EOS through vector-mediated self-repulsion and plays a crucial role in determining the structure of two-fluid stars. In this study, we vary $g_{\chi}/m_{v}$ within the range $0.01$–$0.05$ MeV$^{-1}$, which serves as a representative window motivated by previous analyses~\cite{kumar2025mmc}. While the precise upper bound depends on the choice of the nuclear matter EOS and central energy density ratio of nuclear and DM components, values in this interval remain broadly consistent with neutron star observational data and self-interaction limits inferred from large-scale structure. This parameter space enables us to systematically probe how changes in DM microphysics manifest in equilibrium structure, radial stability, and the emergence of ultra-dense or twin-star configurations.


\section{Radial oscillations and stability analysis}
\label{sec:3}
\subsection{Mirror DM models}
\label{sec:3A}
We first present the two-dimensional frequency landscape of fundamental radial oscillations in two-fluid neutron stars composed of ordinary nuclear matter and mirror DM. As described in Sec.~\ref{sec:2A}, each equilibrium configuration is characterized by a pair of central energy densities, $({\cal E}_{c}^{\rm NM},\ {\cal E}_{c}^{\rm DM})$, and the corresponding fundamental radial oscillation mode is obtained by solving the coupled eigenvalue problem for small-amplitude perturbations. The eigenfrequency, $\omega$, determines whether the system undergoes stable oscillations or is dynamically unstable. In order to represent this in a more physically intuitive form, we define the real-valued characteristic frequency as
\begin{equation}
   f_{\xi} = \sqrt{\omega^2}\ /\ 2\pi,
\end{equation}
which reduces to the standard oscillation frequency (in kHz) when $\omega^2 > 0$, and becomes purely imaginary for unstable configurations with $\omega^2 < 0$.

\begin{figure}[tbp]
    \centering
    \includegraphics[width=\columnwidth]{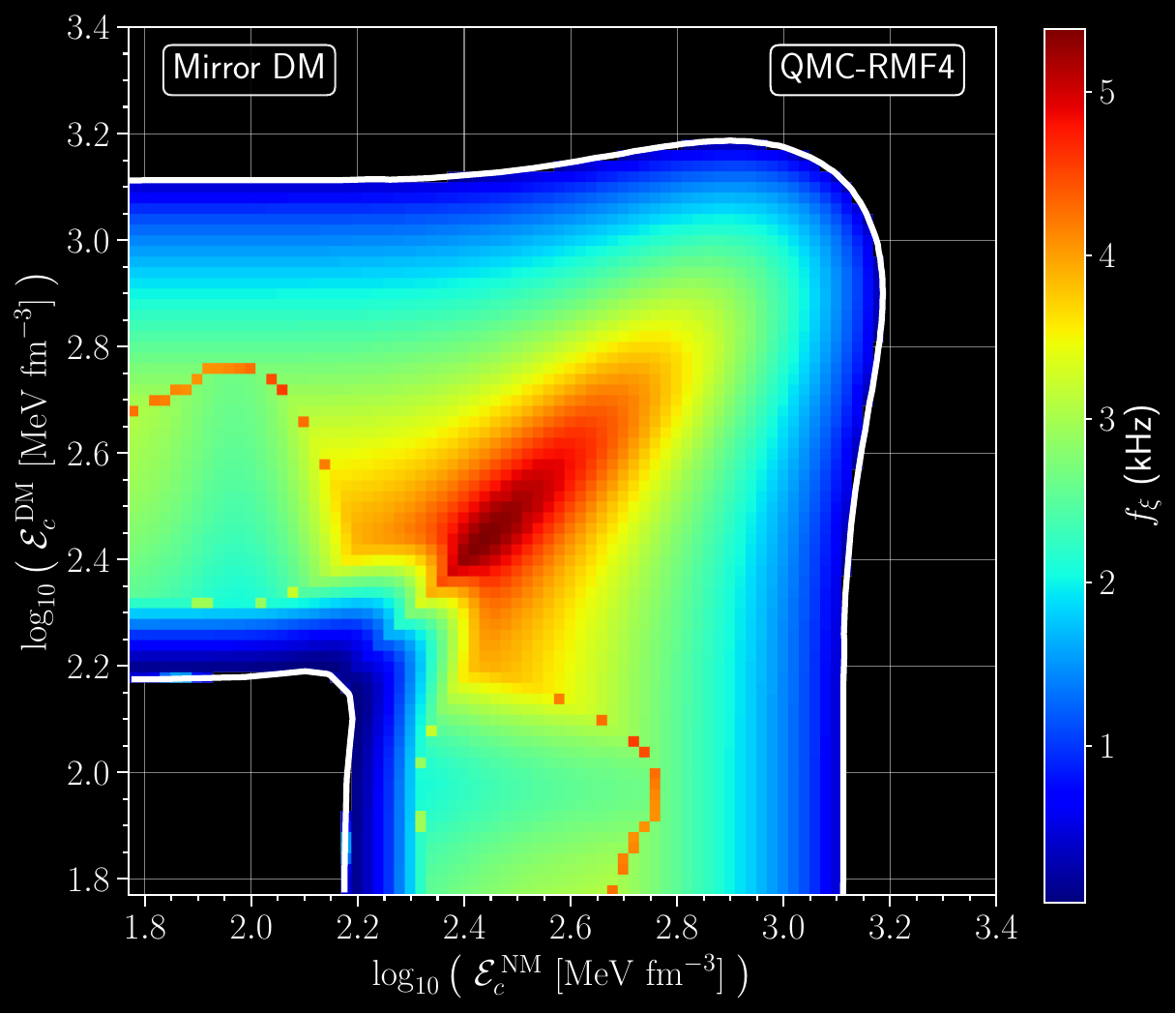} 
    \caption{Fundamental radial mode frequencies $f_{\xi}$ (in kHz) for two-fluid neutron star configurations constructed using the QMC-RMF4 EOS. The 2D parameter space spans central energy densities of nuclear matter (${\cal E}_{c}^{\rm{NM}}$) and mirror DM (${\cal E}_{c}^{\rm{DM}}$), plotted on logarithmic axes. The colormap encodes the value of the fundamental mode frequency $f_{\xi}$, computed for each equilibrium configuration. Solid white curves mark the dynamical stability boundary, where $f_{\xi} = 0$. Configurations enclosed within these contours are dynamically stable, while those outside are unstable to radial perturbations.}
    \label{fig:figure1}
\end{figure}

\begin{figure*}[tbp]
    \centering
    \includegraphics[width=\textwidth]{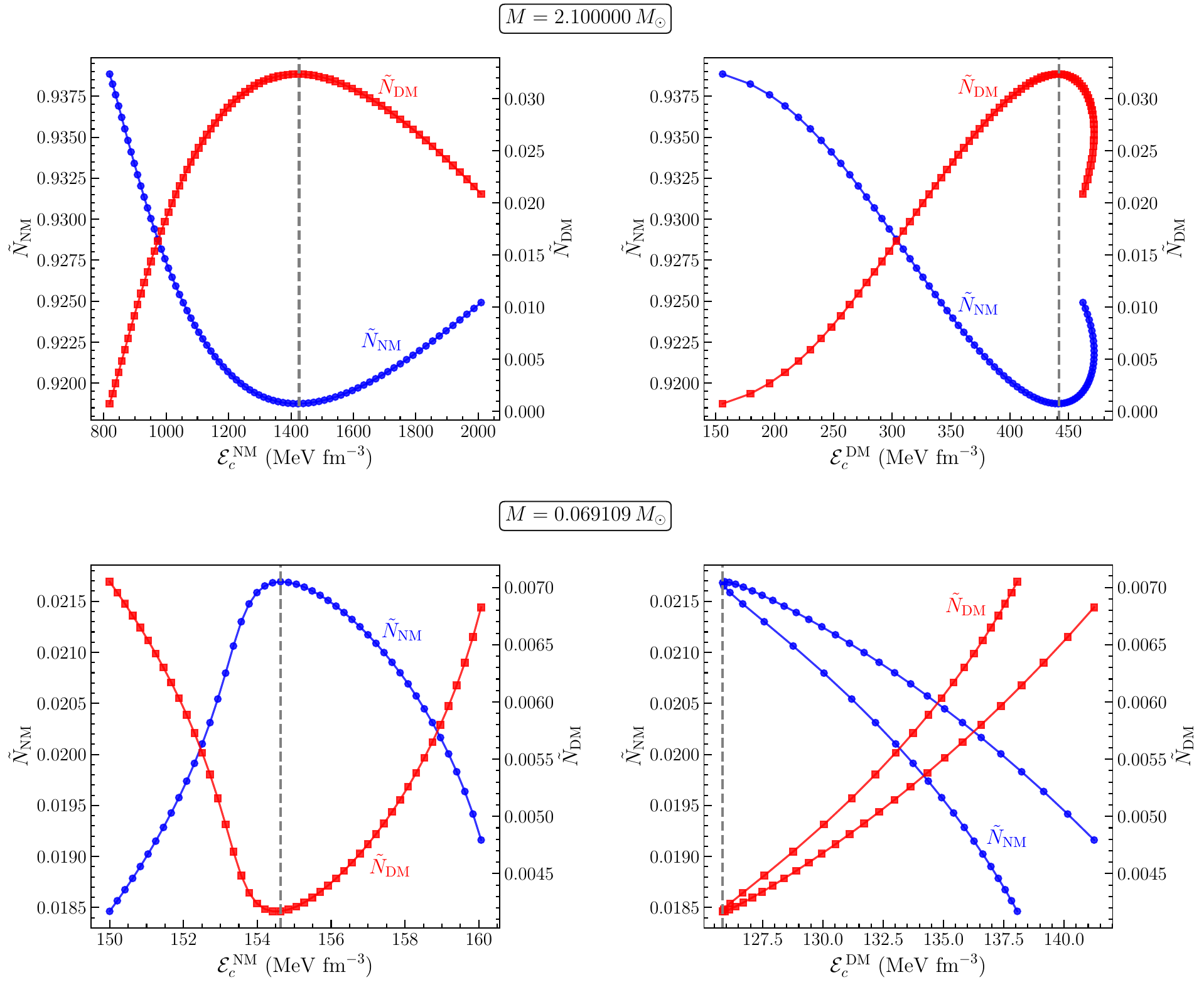  } 
    \caption{Normalized particle numbers for nuclear matter ($\tilde{N}_{\rm{NM}} = N_{\rm{NM}}/N_{\rm max}$) and mirror DM ($\tilde{N}_{\rm{DM}} = N_{\rm{DM}}/N_{\rm max}$) for fixed-mass two-fluid neutron star configuration constructed with the QMC-RMF4 EOS under the mirror DM assumption, plotted as functions of central energy densities \( {\cal E}_{c}^{\rm NM} \)  (left panels) and \( {\cal E}_{c}^{\rm DM} \) (right panels). Here, $N_{\rm max}$ denotes the particle number corresponding to the maximum-mass configuration of a single-fluid star described by the same EOS, and the circles and squares correspond to $\tilde{N}_{\rm{NM}}$ and $\tilde{N}_{\rm{DM}}$ respectively. The top row corresponds to fixed total mass $M = 2.100000\, M_\odot$ and the turning points---where both $\tilde{N}_{\rm{NM}}$ and $\tilde{N}_{\rm{DM}}$ exhibit extrema simultaneously---mark the high-density edge of the dynamical stability domain. The bottom row corresponds to the results with a similar analysis for fixed mass $M = 0.069109\, M_\odot$, revealing the low-density boundary, where both particle numbers again show simultaneous extrema. In each panel, the vertical dashed line denotes the central energy density at which the particle numbers become extrema.}
    \label{fig:figure2}
\end{figure*}

Figure~\ref{fig:figure1} shows a contour map of the fundamental mode frequency $f_{\xi}$ (in kHz) across the two-dimensional parameter space of central energy densities for nuclear matter (${\cal E}_{c}^{\rm NM}$) and mirror DM (${\cal E}_{c}^{\rm DM}$), both in units of MeV/fm$^3$ and on logarithmic scale. The underlying equilibrium configurations are constructed using the QMC-RMF4 EOS, applied identically to both the nuclear and mirror components. This setup corresponds to a minimal mirror DM scenario with symmetric microphysics but independently varying central densities. Only stable configurations with real and positive $f_{\xi}$ are shown; unstable configurations with $\omega^2 < 0$ (i.e., imaginary $f_{\xi}$) are excluded from the plot. The solid white lines denote the dynamical stability boundary, defined by $f_{\xi} = 0$. These contours enclose the stable domain, within which all computed configurations exhibit oscillatory behavior in response to radial perturbations, i.e., $\omega^2>0$.
\begin{figure}[tbp]
    \centering
    \includegraphics[width=\columnwidth]{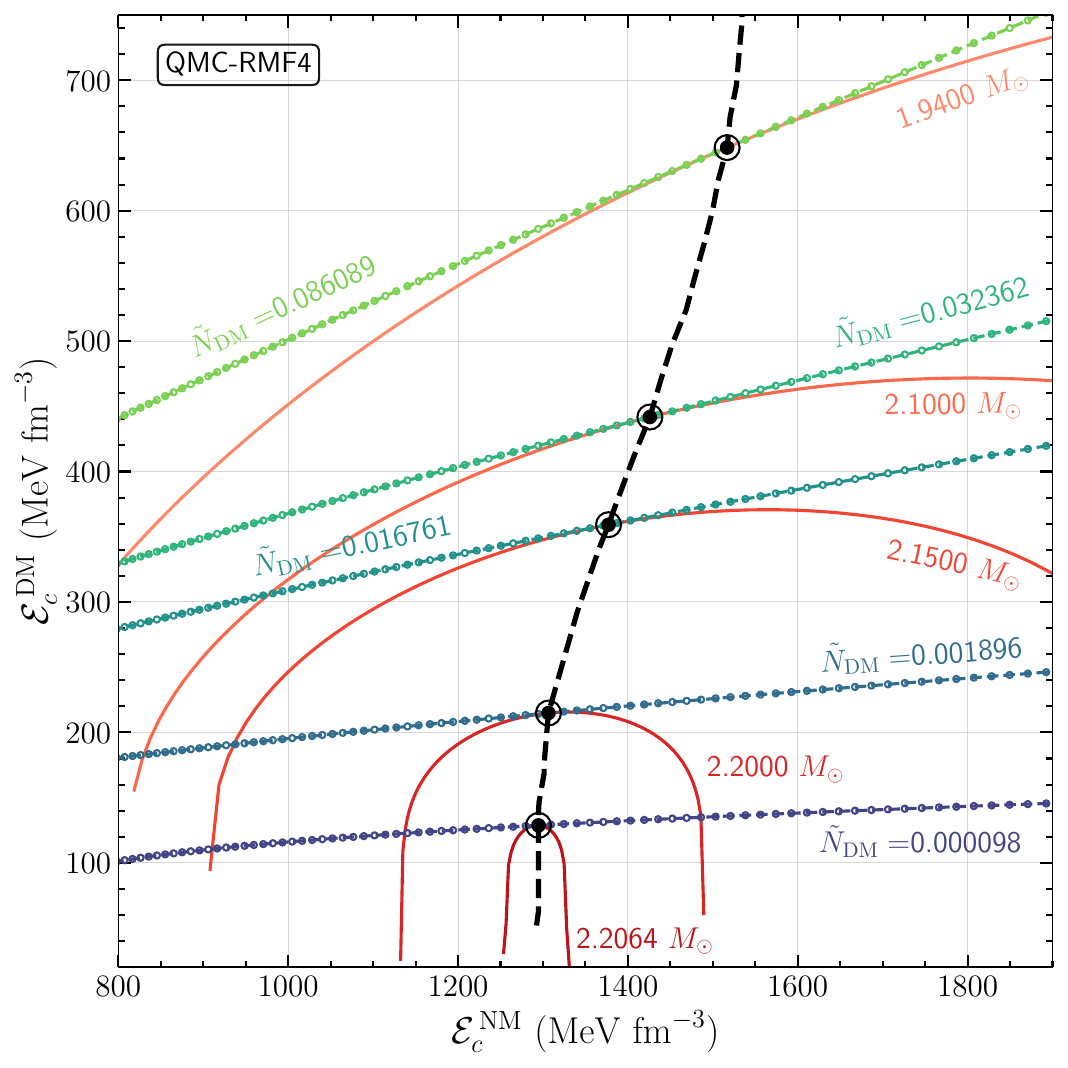} 
    \caption{Contours of fixed total gravitational mass with the solid lines and fixed normalized mirror DM particle number ($\tilde{N}_{\rm DM} = N_{\rm DM}/N_{\rm max}$) in the parameter space of central energy densities for nuclear matter (${\cal E}_{c}^{\rm NM}$) and DM (${\cal E}_{c}^{\rm DM}$), both in units of MeV/fm$^3$ and shown on linear axes. The EOS used for both fluids is QMC-RMF4, corresponding to the mirror DM scenario. The contact points between fixed mass and $\tilde{N}_{\rm DM}$ contours---indicated by bullseye markers and joined by the dashed black line---represent dynamical stability boundary points as determined by the generalized turning-point method.}
    \label{fig:figure3}
\end{figure}

The stable domain exhibits a well-defined island of stability in the central energy density space, enclosed between two instability boundaries: one at low central energy densities (toward the bottom-left), and another at high central densities (toward the top-right). This structure is consistent with expectations from single-fluid neutron stars, which also possess both upper and lower limits of stability thresholds, but here emerges in a genuinely two-dimensional setting. The frequency distribution is symmetric about the diagonal line, reflecting the fact that both fluids are governed by the same microphysical EOS. Since the mirror DM is assumed to mirror the pressure-energy density relation of nuclear matter, the stability and stiffness of the star are equally influenced by either component.

Within the stable region, the values of $f_{\xi}$ reach up to $\sim$ 5.36 kHz, peaking near the center of the stable island, where both fluids contribute comparably to the pressure support. This enhancement arises from the combined stiffness of both components---leading to a greater restoring force against radial perturbations. Toward the boundaries, except for the direction along the diagonal line, as either fluid becomes dominant or diluted, the coupling weakens, and the frequency drops to zero eventually, marking the onset of dynamical instability. In addition to the central region, we also observe narrow, elongated ridges of higher frequency extending horizontally and vertically from the peak, tracing toward the high ${\cal E}_{c}^{\rm NM}$ and high ${\cal E}_{c}^{\rm DM}$ axes, respectively. These ridges correspond to configurations, where one fluid---either nuclear or DM---dominates the equilibrium structure, while the other remains dilute. In such asymmetric setups, the star behaves effectively like a single-fluid object embedded in a weakly coupled background. The relatively high frequencies in these regions arise from the dominant fluid’s compactness and stiffness, which provide strong restoring forces against radial perturbations even in the absence of balanced two-fluid support.

Interestingly, in the high-density regime near the upper-right corner of the stable parameter space, the stability boundary line extends beyond the central energy density, at which a single-fluid QMC-RMF4 star becomes unstable (around ${\cal E}_c \approx 1290\ \mathrm{MeV/fm^3}$). This suggests that the presence of a second fluid component allows the two-fluid star to remain dynamically stable at marginally higher central densities than the one-fluid case. This subtle extension reflects the additional pressure support offered by the second fluid and highlights the need to model such two-fluid stars. We emphasize, however, that this extension refers to the central density domain, not necessarily to the total mass, as two-fluid turning points do not always coincide with a mass extremum.

Finally, we note that a previous study by Ben Kain~\cite{PhysRevD.103.043009} analyzed radial oscillation frequencies in a two-fluid setup with mirror DM, using the SLy4 EOS. While that work focused exclusively on high-density configurations and did not capture the full structure of the stability domain---particularly the low-density boundary---our results, i.e., the frequency structure, are consistent with their findings in the overlapping regime and extend the analysis to uncover the complete stability landscape.

To further examine the dynamical stability of two-fluid stars and validate the boundaries identified in Fig.~\ref{fig:figure1}, we now apply the generalized turning-point method discussed in Sec.~\ref{sec:2B}. This approach relies on identifying equilibrium configurations, where the total gravitational mass, $M$, and the particle numbers of both fluids, $N_{\rm{NM}}$ and $N_{\rm{DM}}$, exhibit simultaneous extrema with respect to variations in the central energy densities. These turning points mark transitions between dynamically stable and unstable configurations, and thus provide an alternative diagnostic of the stability boundary that is independent of the eigenvalue analysis for the radial perturbations.

In Fig.~\ref{fig:figure2}, we illustrate this principle explicitly. The figure shows the normalized particle numbers, $\tilde{N}_{\rm NM} = N_{\rm NM} / N_{\rm max}$ and $\tilde{N}_{\rm DM} = N_{\rm DM} / N_{\rm max}$, evaluated for sequences of two-fluid configurations constructed with fixed total mass and varying central energy densities. The normalization is defined with respect to $N_{\rm{max}}$, which is the particle number corresponding to the maximum-mass configuration of a single-fluid star built from the same EOS (QMC-RMF4). The top row corresponds to fixed total mass $M = 2.100\ M_{\odot}$, which probes the upper edge of the stability domain, while the bottom row displays results for $M = 0.069109\ M_{\odot}$, corresponding to the low-density edge.

\begin{figure}[tbp]
    \centering
    \includegraphics[width=\columnwidth]{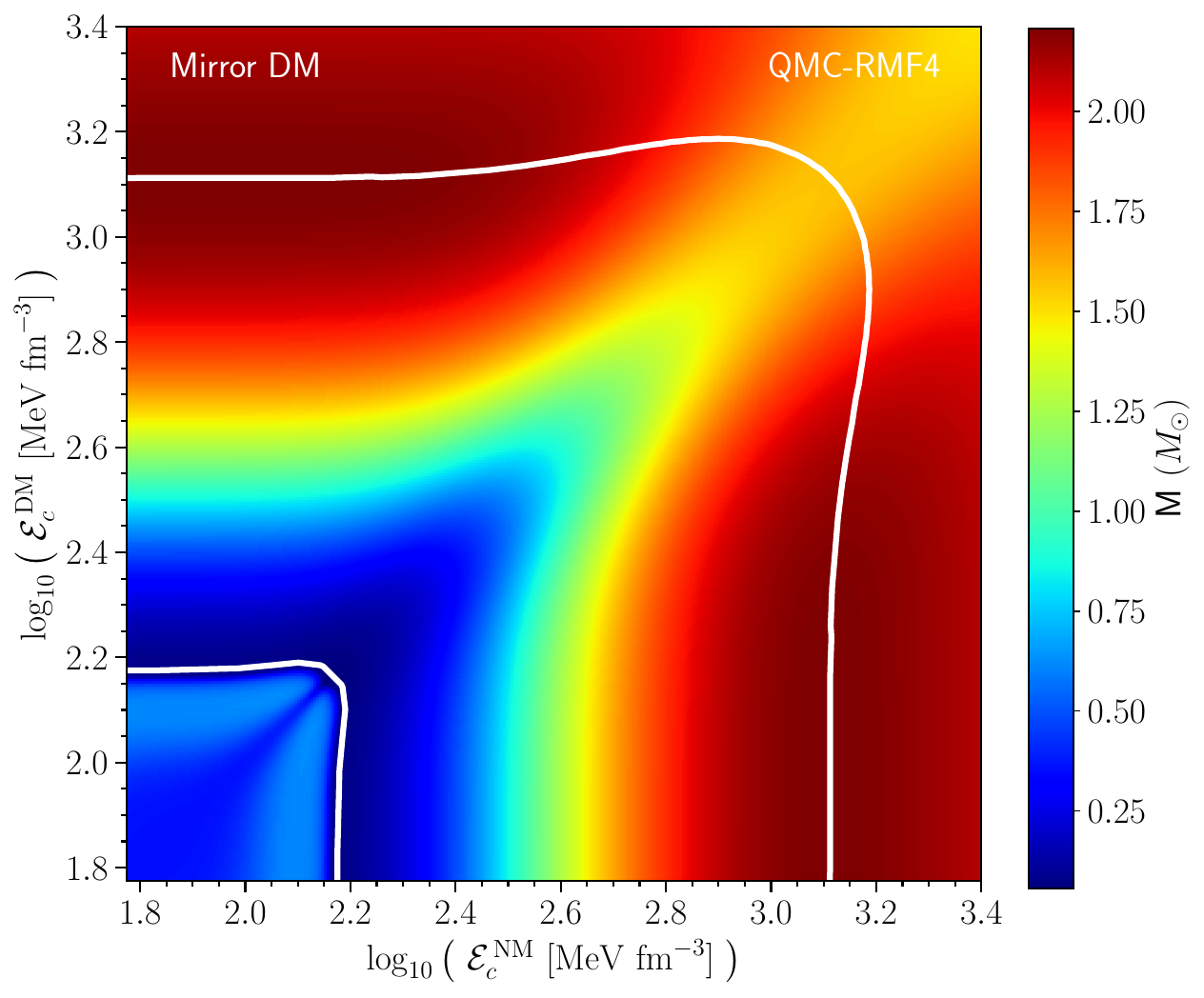} 
    \caption{Contour map of the total gravitational mass, $M$, in units of $M_{\odot}$ across the two-dimensional parameter space of central energy densities for nuclear matter (${\cal E}_{c}^{\rm NM}$) and mirror DM (${\cal E}_{c}^{\rm DM}$), using the QMC-RMF4 EOS for both fluids. The color scale represents the value of $M$, while the overlaid white solid curves trace the full dynamical stability boundary as determined from the generalized turning-point method. Both the high-density (right-edge) and low-density (left-edge) boundaries are shown as white curves, capturing the complete stability domain in the (${\cal E}_{c}^{\rm NM}$, ${\cal E}_{c}^{\rm DM}$) space.}
    \label{fig:figure4}
\end{figure}

For both mass values, the particle numbers are shown as functions of the central energy density of one fluid, while ensuring that the total mass of the star remains constant by tuning the central energy density of the other fluid. The left and right columns correspond to slices along varying ${\cal E}_{c}^{\rm NM}$ and ${\cal E}_{c}^{\rm DM}$, respectively. In each panel, we observe that $\tilde{N}_{\rm NM}$ and $\tilde{N}_{\rm DM}$ attain local extrema at the same point, indicating a simultaneous turning point in both fluid particle numbers. According to the generalized turning-point criterion, such configurations signal the boundary of central energy densities between stability and instability for a fixed total mass. The top row, corresponding to fixed $M = 2.100\ M_{\odot}$, identifies the turning point in the high-density region, while the bottom row reveals the low-density boundary, which remains unexplored in the previous studies. This demonstrates that the generalized turning-point method is sensitive to both ends of the stability domain and reinforces the existence of a two-sided boundary in the (${\cal E}_{c}^{\rm NM}, {\cal E}_{c}^{\rm DM}$) parameter space. We note, however, that tracing the low-density (bottom-left edge) boundary requires significantly higher resolution and numerical precision: turning points in this regime emerge only when mass contours are computed with fine granularity---typically requiring specification to several decimal places, as seen in the use of $M = 0.069109\ M_\odot$ in Fig.~\ref{fig:figure2}. This added complexity highlights the numerical challenges in applying the turning-point method near the low-mass end.

We further illustrate this method geometrically in Fig.~\ref{fig:figure3}, which shows contour lines of fixed gravitational mass (solid red curves) and fixed normalized DM particle number $\tilde{N}_{\rm DM}$ (dashed green curves) in the two-dimensional space of central energy densities. According to the generalized turning-point criterion as discussed in Sec.~\ref{sec:2B}, the stability boundary occurs at points where the contour families---associated with $M$, $N_{\rm{NM}}$, and $N_{\rm{DM}}$---become mutually tangent in the parameter space. In the present plot, only the $\tilde{N}_{\rm DM}$ contours are shown alongside the mass contours, while the $\tilde{N}_{\rm NM}$ curves are not displayed explicitly, as their trajectory coincides closely with the mass contours in this mirror matter setup. The bullseye markers indicate the locations where a fixed-mass contour becomes tangent to a fixed $\tilde{N}_{\rm DM}$ contour. These points, connected by a dashed black line, collectively trace the dynamical stability boundary predicted by the turning-point method.

To consolidate the results of the turning-point analysis, in Fig.~\ref{fig:figure4}, we present a global contour map of the total gravitational mass, $M$, across the full two-dimensional space of central energy densities. The mass values are shown in units of $M_{\odot}$, with the colormap representing the continuous variation in stellar mass across configurations. Superimposed on this map is the complete dynamical stability boundary (solid white curve), derived using the generalized turning-point method described above. The boundary is constructed from turning points identified via tangency conditions involving the contours of the mass and particle number, as illustrated in Figs.~\ref{fig:figure2} and~\ref{fig:figure3}. By visualizing the global mass landscape alongside the turning-point stability curve, this figure provides a comprehensive overview of the stable parameter space allowed for two-fluid stars in the mirror DM scenario.

Notably, the curve formed by these points coincides with the boundary obtained by solving the eigenvalue problem for the radial perturbations (Fig.~\ref{fig:figure1}). A quantitative comparison reveals that the two methods agree to within 1\% accuracy across the entire range of central energy densities, with any discrepancies lying entirely within numerical resolution. This excellent agreement confirms the consistency between the dynamical (eigenfrequency) and thermodynamic (turning-point) criteria for stability in two-fluid stars. It not only validates the robustness of both approaches but also highlights the practical utility of the generalized turning-point method for efficiently mapping stability boundaries, particularly in high-dimensional parameter spaces where full oscillation analysis is computationally expensive.

In this section, we discuss the stability within the mirror DM model, adopting only QMC-RMF4 EOS, but one can discuss qualitatively the same features even with a different EOS. For reference, we show the behavior with different EOS, i.e., DD2 and QHC21-BT, in Appendix~\ref{sec:appendx1}.

\subsection{Self-interacting fermionic DM models}
\label{sec:3B}
\begin{figure*}[tbp]
    \centering
    \includegraphics[width=\textwidth]{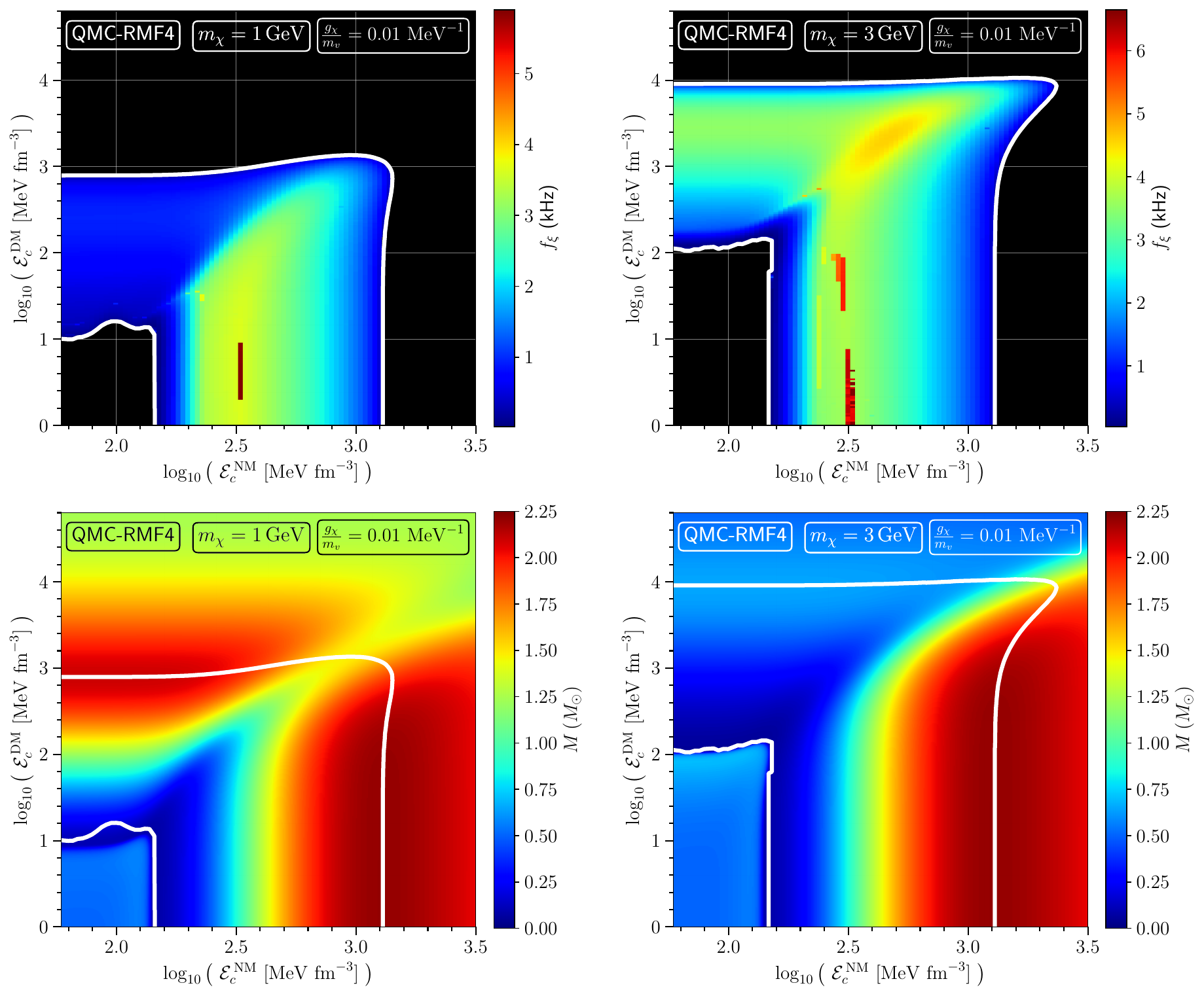} 
    \caption{Top row: Contour maps of the fundamental radial oscillation frequency $f_{\xi}$ (in kHz) for two-fluid neutron star configurations constructed using the QMC-RMF4 nuclear EOS and a self-interacting DM model with coupling ratio $g_\chi/m_v = 0.01\ \mathrm{MeV}^{-1}$, shown for two DM particle masses: \( m_\chi = 1\ \mathrm{GeV} \) (left) and \( m_\chi = 3\ \mathrm{GeV} \) (right). Bottom row: Corresponding maps of the total gravitational mass \( M \) (in \( M_{\odot} \)) over the same central energy density parameter space. In all panels, white curves indicate the dynamical stability boundaries. The high-density (right-edge) boundary is determined independently through the eigenvalue analysis from the radial oscillations (top) and the generalized turning-point method (bottom); while the low-density (left-edge) boundary---shown here for completeness---is derived from the vanishing of the oscillation frequency, i.e., $f_\xi = 0$, due to the increased numerical complexity of resolving the turning-point surface in this regime.}
    \label{fig:figure5}
\end{figure*}

To explore the influence of microphysical DM properties on the stability of two-fluid neutron stars, we now turn to the case of fermionic DM with vector-mediated self-interactions. In this scenario, the DM fluid is modeled as a cold self-interacting Fermi gas, as described in Sec.~\ref{sec:2C}, with its EOS depending on both the DM particle mass, $m_{\chi}$, and the vector coupling ratio, $g_{\chi}/m_{v}$. The nuclear component is modeled using QMC-RMF4 EOS, as before.

Figure~\ref{fig:figure5} presents the results for two representative vector DM models: one with $m_{\chi} = 1$ GeV (left column) and the other with $m_{\chi} = 3$ GeV (right column). In both cases, the coupling ratio is fixed at $g_{\chi}/m_{v} = 0.01$ MeV$^{-1}$. The top row shows the fundamental radial oscillation frequency, $f_{\xi}$, while the bottom row shows the total gravitational mass, $M$, both mapped over the two-dimensional space of the central energy densities, ${\cal E}_{c}^{\rm NM}$ and ${\cal E}_{c}^{\rm DM}$. The white contours in each panel mark the boundaries of the dynamically stable region. The high-density (right-edge) boundaries are derived independently of the radial oscillation and generalized turning-point methods (top and bottom panels, respectively), while the low-density (left-edge) boundaries are extracted solely from the oscillation analysis, due to the increased numerical difficulty in resolving the turning-point surface near the low-mass limit. As discussed previously, locating these low-density turning points using the generalized method requires extremely fine resolution in mass and particle numbers, making the eigenmode analysis a more practical and robust tool for identifying the left-edge stability boundary in such regimes.

With these frequency and mass maps in hand, we now examine how the structure of the stability region responds to changes in the DM microphysics. First, we again observe a two-sided stability boundary enclosing a finite region of parameter space, showing that the presence of self-interacting DM leads to both high- and low-density instabilities, analogous to the mirror matter case. However, the size and shape of the stability region change significantly as DM particle mass increases from 1 to 3 GeV, as DM EOS becomes progressively softer. To maintain hydrostatic balance in this case, stable configurations require significantly higher central energy densities in the dark sector. This trend is evident from the position of the stability boundaries along the vertical axis: for $m_{\chi} = 1$ GeV, the lower and upper boundary edges intersect the DM axis at approximately $10$ MeV/fm$^3$ and $800$ MeV/fm$^3$, respectively, whereas for $m_{\chi} = 3$ GeV, these intersections shift to roughly $100$ MeV/fm$^3$ and $8000$ MeV/fm$^3$. These shifts reflect the increased gravitational compression required to stabilize stars with a softer DM component. Moreover, the stability boundary on right-edge shifts noticeably toward higher nuclear matter densities in the case with $m_{\chi} = 3$ GeV, indicating that heavier DM particle allows stable configurations with more dense nuclear cores. This trend is also reflected in the deformation of the mass contour structures in the lower panels, where the boundary extends deeper into the high ${\cal E}_{c}^{\rm{NM}}$ regime. 

\begin{figure*}[tbp]
    \centering
    \includegraphics[width=\textwidth]{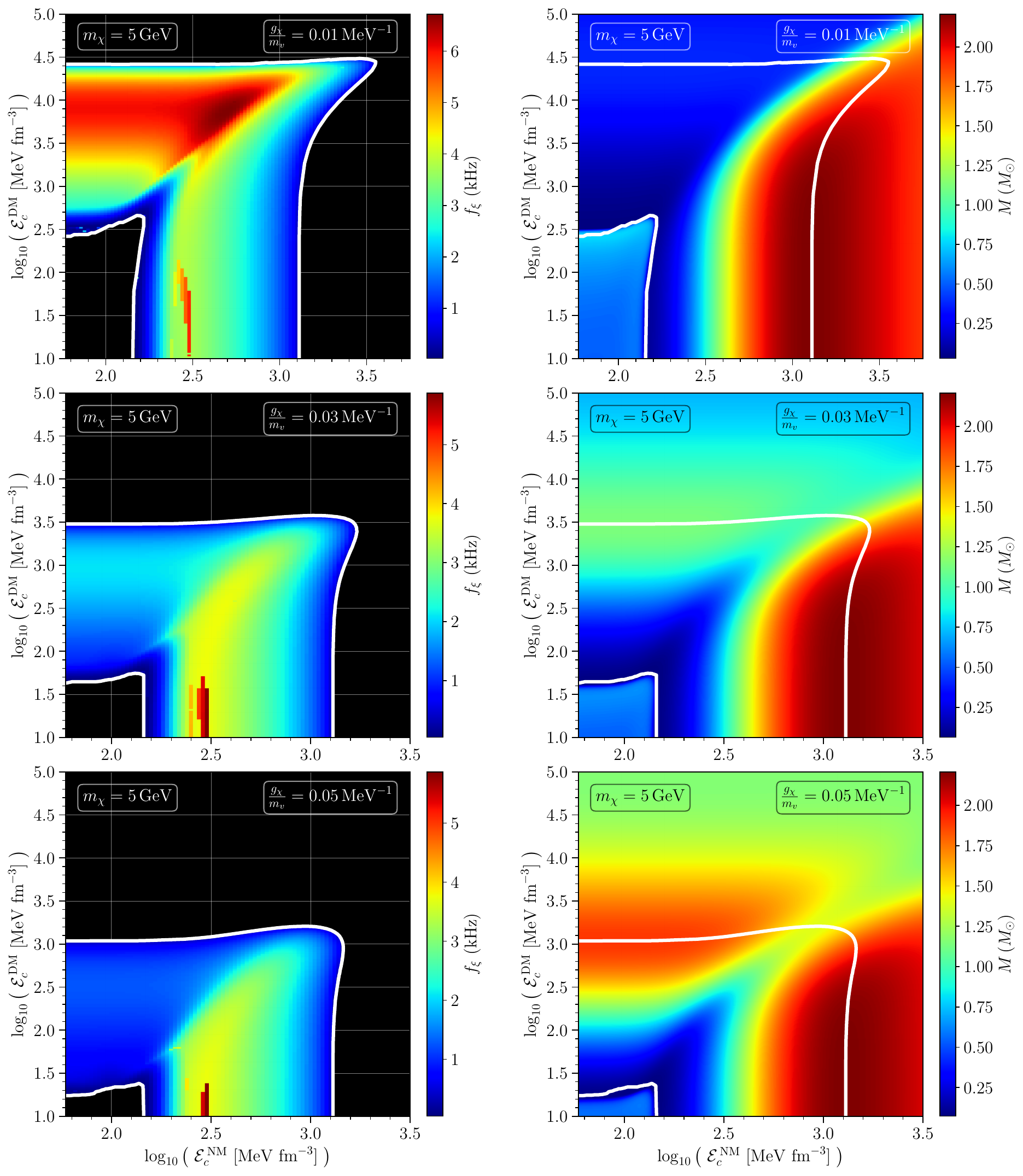} 
    \caption{Stability analysis of two-fluid neutron star configurations with fixed DM particle mass, i.e., $m_\chi = 5\ \mathrm{GeV}$, constructed using the QMC-RMF4 EOS for the nuclear component and a vector-interacting fermionic DM model. The left column is the fundamental radial mode frequencies $f_{\xi}$ (in kHz) shown as 2D contour maps over the central energy densities of nuclear matter (${\cal E}_{c}^{\rm NM}$) and DM (${\cal E}_{c}^{\rm DM}$), for three values of the DM self-interaction parameter: $g_\chi/m_v$ = 0.01 MeV$^{-1}$ (top), 0.03 MeV$^{-1}$ (middle), and 0.05 MeV$^{-1}$ (bottom). The right column is the corresponding total gravitational mass $M$ (in $M_{\odot}$) across the same parameter space. In each panel, the white contour denotes the dynamical stability boundary. The high-density (right-edge) boundaries are independently derived from the oscillation frequency analysis (left column) and the generalized turning-point method (right column), while the low-density (left-edge) boundaries are obtained from the vanishing frequency condition, i.e., $f_{\xi} = 0$, with the eigen mode analysis.}
    \label{fig:figure6}
\end{figure*}

These differences in DM EOS also manifest in the overall gravitational mass structure of the configurations. For $m_{\chi} = 1$ GeV, the mass contours in the high central energy DM region---toward the upper stability boundary corresponding to the dark halo structures---rise more steeply and produce higher maximum masses compared to the case with $m_{\chi} = 3$ GeV. This is consistent with the stiffer EOS for lower $m_{\chi}$, which provides greater pressure support at high densities. In contrast, the $m_{\chi} = 3$ GeV model, characterized by a softer DM EOS, yields noticeably lower masses in the same region due to reduced pressure support from the dark sector. These results underscore the impact of DM microphysics on the maximum mass and density structure of two-fluid stars, even when the nuclear component remains fixed.

In the top two panels of Fig.~\ref{fig:figure5}, we observe narrow patches of artificially elevated frequency within the stable region. These features are numerical artifacts arising from localized instabilities during the integration of the perturbation equations—typically caused by steep gradients, interpolation irregularities, or sensitive boundary matching in the background profiles. Although they manifest as spurious spikes in the frequency map, they are confined to small regions and do not influence the determination of the overall stability boundary. Separately, in the top-right panel, we also observe a lower ridge of genuinely enhanced frequency extending toward the ${\cal E}_{c}^{\rm DM} \rightarrow 0$ limit, particularly around $\log_{10}({\cal E}_{c}^{\rm NM}) \sim 2.5$. This structure corresponds to configurations where the nuclear matter core alone supports the star, with the dark matter component remaining dilute. These effectively single-fluid nuclear matter stars exhibit higher oscillation frequencies due to their compactness and stiffness, resulting in a physically meaningful ridge of elevated $f_\xi$.

To isolate the effect of vector coupling strength on stability, we next examine configurations with a fixed DM particle mass of $m_\chi = 5\ \mathrm{GeV}$, while varying the coupling ratio $g_{\chi}/m_{v}$ across three representative values: 0.01, 0.03, and 0.05 MeV$^{-1}$. Fig.~\ref{fig:figure6} summarizes the results for this scenario using the QMC-RMF4 EOS for the nuclear component. The left column displays contour maps of the fundamental radial oscillation frequency, $f_{\xi}$, while the right column shows the corresponding total gravitational mass distributions. In each panel, the high-density (right-edge) stability boundary is determined independently via the oscillation analysis and the generalized turning-point method. The two approaches yield virtually identical high-density boundary curves for all coupling values, underscoring not only their mutual consistency but also the reliability of each method as an independent diagnostic of dynamical stability. Meanwhile, the low-density (left-edge) boundaries are traced from the vanishing frequency condition, i.e., $f_{\xi} = 0$, due to numerical limitations of the turning-point approach in this regime, as discussed earlier.

The variation of the $g_{\chi}/m_{v}$ parameter directly affects the stiffness of the DM EOS. At fixed $m_{\chi}$, increasing $g_{\chi}/m_{v}$ leads to stronger self-repulsion among DM particles, resulting in a stiffer EOS. This trend is evident in the structure of the stability domain. For the weakest coupling, $g_{\chi}/m_{v} = 0.01$ MeV$^{-1}$, the DM fluid remains soft, and stable configurations extend into extremely high central densities for both nuclear and dark components. In particular, the right-edge boundary at $g_{\chi}/m_{v} = 0.01$ MeV$^{-1}$ reaches significantly deeper into the dense nuclear matter regime---about $\rm{log}_{10}({\cal E}_{c}^{\rm NM}) \approx 3.55$ \footnote{Specifically, the stability boundary for $g_{\chi}/m_{v} = 0.01$ reaches ${\cal E}_{c}^{\rm NM} \approx 3560$ MeV/fm$^3$, which is approximately 2.76 times higher than the instability threshold for a single-fluid neutron star constructed with QMC-RMF4 model, located at $\approx 1290$ MeV/fm$^3$.}---reflecting the increased gravitational compression required to stabilize such configurations.

As the coupling ratio increases to $g_{\chi}/m_{v} = 0.03$ and $0.05$ MeV$^{-1}$, the DM EOS becomes progressively stiffer, and the overall stability region shifts downward and to the left in the central energy density plane. This shift reflects the fact that a stiffer DM model can provide sufficient pressure support even at lower densities, allowing equilibrium to be maintained by counterbalancing the effects of extreme gravitational compression. Accordingly, the right-edge boundary retreats toward more moderate nuclear matter densities---closer to the instability threshold of a pure nuclear matter star described by the QMC-RMF4 EOS. Notably, in these stiffer cases, the gravitational mass increases significantly in the high ${\cal E}_{c}^{\rm DM}$ and low ${\cal E}_{c}^{\rm NM}$ corner of the parameter space. This behavior stems from the enhanced pressure support of the dark sector, which enables the formation of massive configurations even when the nuclear component is dilute. These high-mass, DM-dominated stars likely exhibit extended halos in which the dark fluid envelops a comparatively compact nuclear core, highlighting how DM self-interactions reshape both the mass distribution and the internal structure of two-fluid stars.

\begin{figure}[tbp]
    \centering
    \includegraphics[width=\columnwidth]{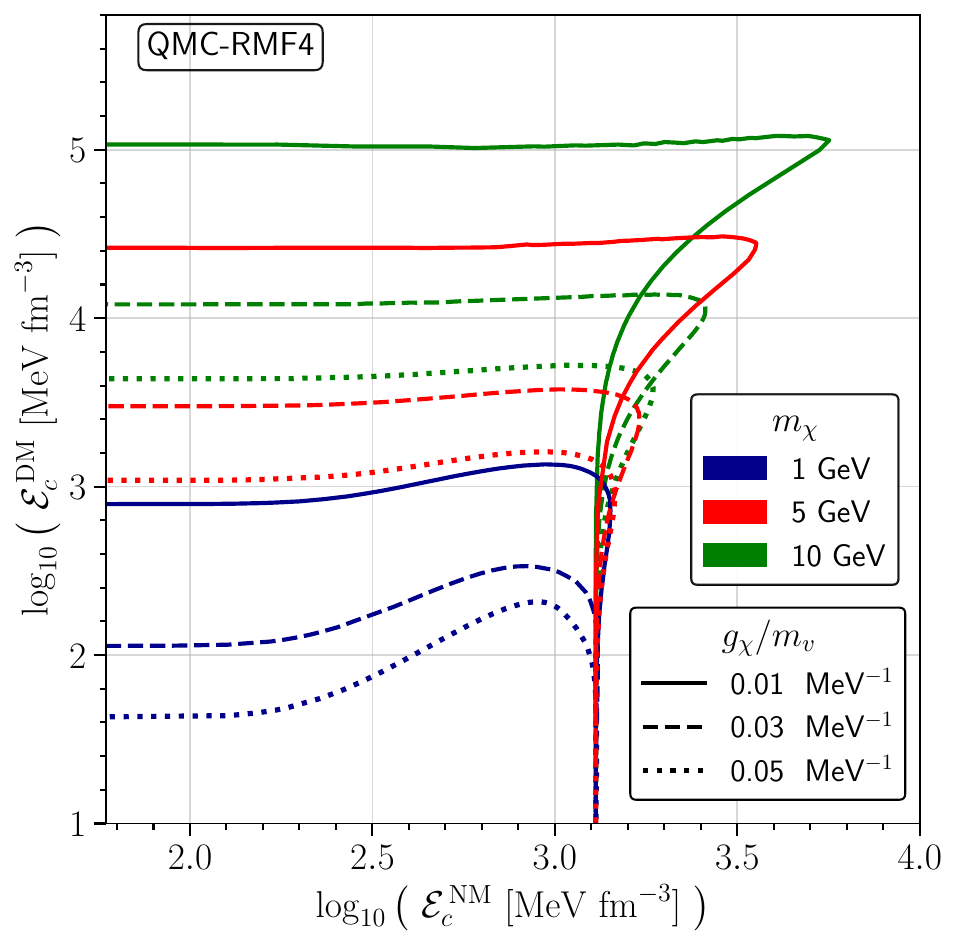} 
    \caption{Comparison of high-density (right-edge) dynamical stability boundaries for two-fluid neutron stars with vector-interacting DM, using the QMC-RMF4 nuclear EOS. The curves represent different combinations of DM particle mass $m_{\chi} = 1, 5, 10$ GeV and coupling strengths $g_{\chi}/m_{v} = 0.01, 0.03, 0.05$ MeV$^{-1}$. Colors distinguish the DM masses (dark blue: 1 GeV, red: 5 GeV, green: 10 GeV), while line styles (solid, dashed, dotted) distinguish the coupling strengths. The boundary curves trace the upper edge of the dynamically stable region in the $\rm{log}_{10}({\cal E}_{c}^{\rm NM})-\rm{log}_{10}({\cal E}_{c}^{\rm DM})$ space.}
    \label{fig:figure7}
\end{figure}

The oscillation frequency maps also reflect the systematic influence of DM self-coupling. At the lowest coupling, $g_{\chi}/m_{v} = 0.01$ MeV$^{-1}$, the peak values of $f_{\xi}$ are higher, and the stable region extends to very high DM central densities. As the coupling increases to 0.03 and 0.05 MeV$^{-1}$, the frequencies decrease modestly, and the stability domain shifts downward along the ${\cal E}_{c}^{\rm DM}$ axis, consistent with the stiffening of the DM EOS. This trend reflects the fact that stiffer DM requires less central compression to stabilize the configuration, resulting in lower oscillation frequencies due to reduced overall compactness. Additionally, in all three cases, we observe ridges of elevated $f_{\xi}$ extending toward the ${\cal E}_{c}^{\rm DM} \rightarrow 0$ limit---specifically, around $\log_{10}({\cal E}_{c}^{\rm NM}) \approx 2.4-2.5$---indicating configurations, where the nuclear matter component dominates the restoring forces against radial perturbations, while the dark sector remains dilute. These asymmetric stars behave effectively like single-fluid nuclear stars with a subdominant DM component, leading to higher-frequency oscillations driven by the compactness of the nuclear core. 

These results shown in Fig.~\ref{fig:figure6} demonstrate that even for a fixed DM particle mass, the self-interaction strength can significantly influence the structure, stability, and oscillation properties of two-fluid stars. The ability of DM to provide pressure support at different densities is not solely governed by its particle mass, but also by the interaction physics---an important factor when interpreting the internal composition of neutron stars. While radial oscillations themselves are not directly observable via gravitational waves due to their spherical symmetry, their role in defining the dynamical stability boundary has important consequences. The location and shape of this boundary determine the maximum mass and compactness achievable for two-fluid configurations, which in turn affect global observables such as tidal deformability, quasinormal mode spectra, and post-merger remnant stability. Notably, the radial mode frequencies in our models reach several kilohertz---comparable to the typical frequencies of quasinormal modes that do emit gravitational waves from compact objects. In highly dynamical environments, such as merger remnants or magnetar flares, nonlinear coupling between radial and nonradial modes may amplify surface deformations or trigger quasi-periodic emission signatures, potentially leaving imprints in electromagnetic or gravitational wave observations.

To provide a broader perspective on how DM particle mass and interaction strength influence dynamical stability, we compile in Fig.~\ref{fig:figure7} the high-density (right-edge) boundaries obtained from the generalized turning-point method across all models discussed in this study. The figure compares the results for three DM particle masses, i.e., $m_{\chi} = 1, 5, 10$ GeV, and three coupling strengths, i.e., $g_{\chi}/m_{v} = 0.01, 0.03, 0.05$ MeV$^{-1}$, using the QMC-RMF4 nuclear EOS in all cases. For comparison, corresponding results based on the QHC21-BT EOS are shown in Appendix~\ref{sec:appendx2} to demonstrate the robustness of these trends across different nuclear matter models.

The behavior of the stability boundaries across these models reveals a consistent trend: for a given DM particle mass, increasing the coupling $g_{\chi}/m_{v}$ leads to a stiffer EOS, which in turn shifts the stability region toward lower DM central densities. When comparing across different masses, the 
$m_{\chi} = 5$ and $10$ GeV models show a clear extension of the right-edge boundary deeper into the high nuclear matter density regime. This effect is most pronounced for $m_{\chi} = 10$ GeV with $g_{\chi}/m_{v} = 0.01$ MeV$^{-1}$, where the stability boundary extends well beyond the central energy density threshold for a neutron star of single-fluid nuclear matter constructed with the QMC-RMF4 EOS. This reflects the extremely soft nature of the DM EOS in this limit, which requires substantial compression from the nuclear core to maintain equilibrium.

In contrast, for $m_{\chi} = 1$ GeV, the right-edge boundary bends upward in the ${\cal E}_{c}^{\rm DM}$ direction, especially at higher couplings ($g_{\chi}/m_{v} = 0.03$ and $0.05$ MeV$^{-1}$). This behavior reflects a transition to DM-dominated configurations, where the dark fluid becomes the primary contributor to the hydrostatic support, and the nuclear component plays a comparatively subdominant role. Notably, the stability boundaries for the $m_{\chi} = 1$ GeV case remain largely confined to nuclear central densities near the instability threshold of the corresponding single-fluid neutron star with QMC-RMF4 EOS.

Finally, all boundary curves converge toward a common narrow region at low DM central densities---along the ${\cal E}_{c}^{\rm NM}$ axis---regardless of the DM particle mass or coupling. This coalescence illustrates that in the limit, where the DM becomes dilute, the structure and stability of the star are governed almost entirely by the nuclear matter, and the influence of the dark sector becomes negligible. Taken together, these features underscore how the shape and orientation of the stability boundary encode the relative dominance of the nuclear and dark components, and how this balance shifts with both $m_{\chi}$ and $g_{\chi}/m_{v}$.

\section{Ultra-dense configurations and twin-star structures}
\label{sec:4}
%
\begin{figure}[tbp]
    \centering
    \includegraphics[width=\columnwidth]{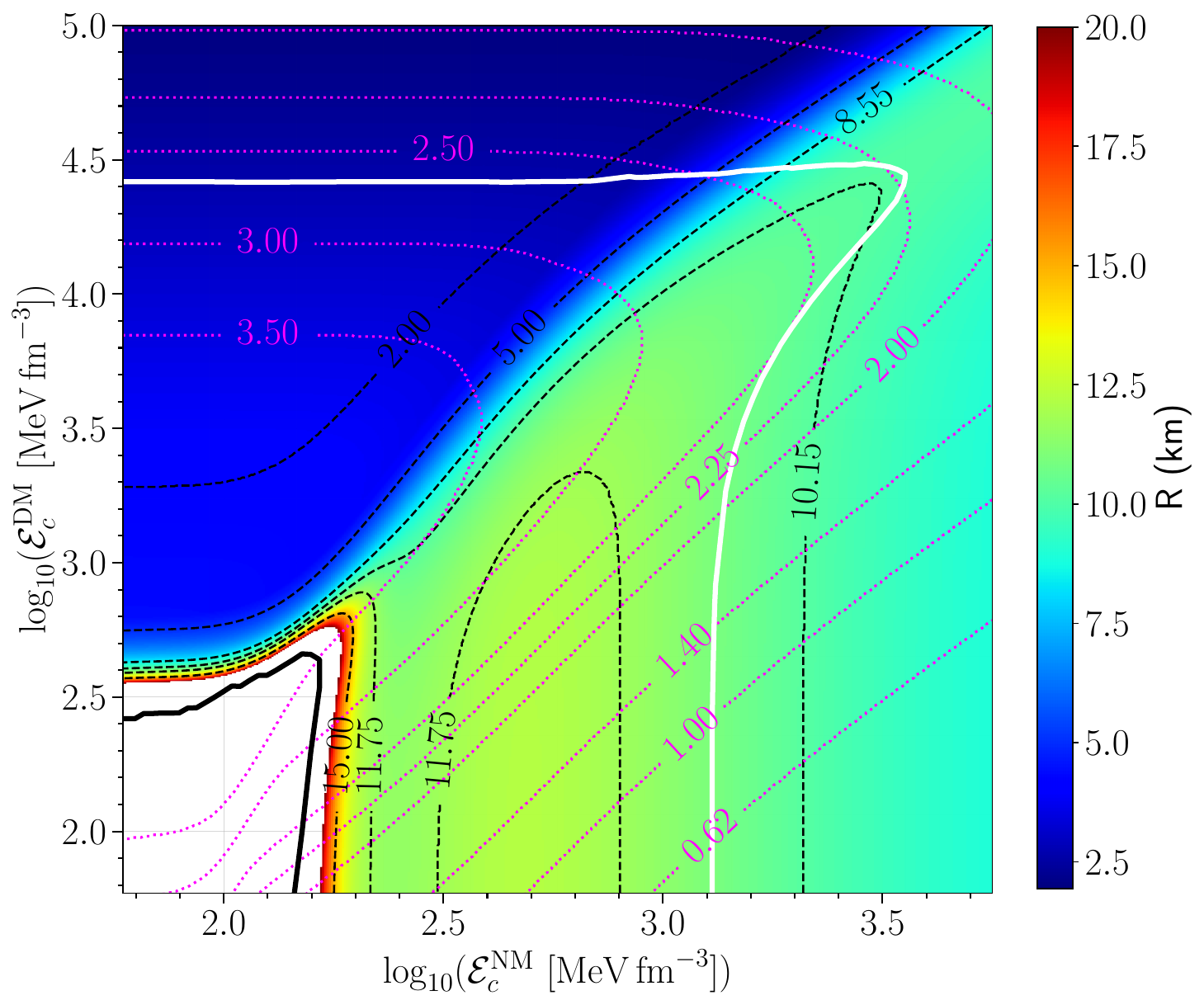}
    \caption{Radius structure of two-fluid neutron stars composed of QMC-RMF4 nuclear matter and vector-interacting DM with  $m_\chi = 5\, \mathrm{GeV}$ and $g_\chi/m_v = 0.01\, \mathrm{MeV}^{-1}$. The colorbar shows the overall stellar radius, $R = \max\ (R_{\rm NM}, R_{\rm DM})$ in kilometers (upto only 20 km), defined as the larger of the two fluid surfaces. Dashed black curves show contours of constant nuclear matter radius \( R_{\rm NM} \), while dotted magenta curves show contours of constant DM radius \( R_{\rm DM} \), both labeled in km. Solid white line marks the high-density stability boundary, and solid black line traces the low-density dynamical stability boundary.} 
    \label{fig:figure8}
\end{figure}

\begin{figure*}[tbp]
    \centering
    \includegraphics[width=\textwidth]{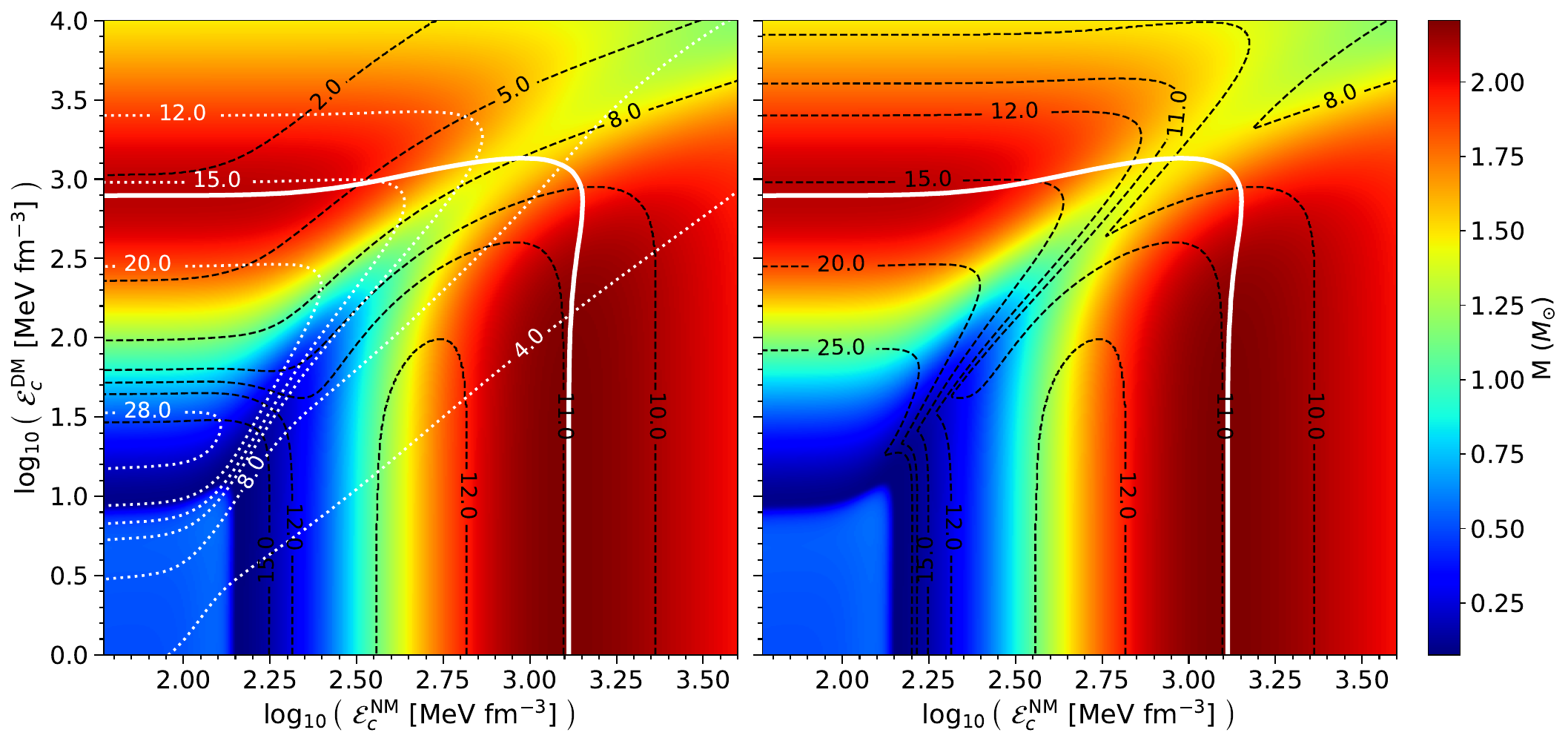}
    \caption{Twin-star configurations in two-fluid neutron stars composed of QMC-RMF4 nuclear matter and vector-interacting DM with $m_{\chi} = 1$ GeV and $g_{\chi}/m_{v} = 0.01$ MeV$^{-1}$. Both panels display contour maps of the total gravitational mass $M$ (in units of $M_{\odot}$) as a function of central energy densities $\log_{10}({\cal E}_{c}^{\rm NM})$ and $\log_{10}({\cal E}_{c}^{\rm DM})$. Left panel: Overlaid dashed black and dotted white contours represent nuclear matter and DM radii (in km), respectively. Right panel: Overlaid contours indicate the overall stellar radius\( R = \max\ (R_{\rm NM}, R_{\rm DM}) \). Solid white lines in both panels mark the high-density dynamical stability boundary. In the vicinity of the boundary, configurations with nearly identical masses but substantially different radii emerge, indicating the presence of twin-star-like solutions supported by two-fluid dynamics.}
    \label{fig:figure9}
\end{figure*}

One of the most striking features of the two-fluid models with vector-interacting DM is the emergence of stable equilibrium configurations that reach extreme nuclear densities while maintaining radii comparable to ordinary neutron stars. These ultra-dense stars appear in the upper-right corner of the stability domain, especially for soft DM EOS such as the case with $m_{\chi} =  5$ GeV and $g_{\chi}/m_{v} = 0.01$ MeV$^{-1}$, shown in Fig.~\ref{fig:figure8}. As previously discussed in Sec.~\ref{sec:3A}, the dynamical stability boundary for this model extends far beyond the instability threshold of a single-fluid nuclear star, reaching up to $\log_{10}({\cal E}_{c}^{\rm NM}) \approx 3.55$, or roughly $3560$ MeV/fm$^{-3}$. This corresponds to central densities more than twice as large as those allowed in isolated single-fluid nuclear matter stars constructed from the QMC-RMF4 EOS, indicating the existence of an exotic class of ultra-compact, two-fluid configurations.

Despite their extraordinary central densities, these configurations exhibit radii in the typical neutron star range---between 9 and 11 km---making them observationally indistinguishable from ordinary neutron stars in terms of surface size. As shown in Fig.~\ref{fig:figure8}, the overall stellar radius, defined as \( R = \max\ (R_{\rm NM}, R_{\rm DM}) \), remains moderate even in the upper-right corner of the parameter space, where \( \log_{10}({\cal E}_{c}^{\rm NM}) \approx 3.55 \) and \( \log_{10}({\cal E}_{c}^{\rm DM}) \approx 4.42 \). The structure of these stars, however, is highly stratified. The nuclear matter radius $R_{\rm{NM}}$, indicated by the dashed black contours, extends to roughly $9-10$ km, while the DM radius $R_{\rm{DM}}$, shown by the dotted magenta lines, remains compact---confined to about $2-2.50$ km. The dark sector in these configurations forms a gravitationally compressed inner core, enveloped by an extended nuclear fluid, i.e., dark core structures, that provides the bulk of the pressure support and defines the stellar surface.

The extreme compactness of the DM core, stabilized deep within an ultra-dense nuclear mantle, points to a highly nontrivial internal composition. The resulting object resembles a hybrid-like configuration, with the potential to harbor exotic phases in both components under such extreme conditions. From an observational perspective, these stars may be virtually indistinguishable from ordinary neutron stars in radius, mass, or tidal deformability, yet their interiors probe densities far beyond those accessible in conventional single-fluid stars. Similar ultra-dense configurations are also found in models with heavier DM---for instance, in the $m_{\chi} = 10$ GeV, $g_{\chi}/m_{v} = 0.01$ MeV$^{-1}$ case---not shown here for brevity but exhibiting analogous stability boundaries extending deep into the high-density regime. In this sense, ultra-dense two-fluid stars expand the landscape of theoretically viable neutron star models, offering a natural astrophysical laboratory for exploring fundamental QCD at supranuclear densities and presenting important opportunities for future multimessenger constraints.

In addition to ultra-dense compact configurations, another intriguing feature arising in the two-fluid landscape is the existence of twin-star structures. These are pairs of stable stellar configurations that share identical gravitational masses but differ significantly in radius and internal composition. In traditional one-fluid neutron star models, such solutions typically arise from strong first-order phase transitions in the EOS. However, in two-fluid systems, twin-like behavior can also emerge dynamically from the coupled equilibrium between nuclear matter and the dark sector. This phenomenon is clearly visible in Fig.~\ref{fig:figure9}, where the color scale represents total gravitational mass $M$ (in units of $M_{\odot}$), overlaid with different radius contours in the two panels. In the left panel, black dashed line contours indicate nuclear matter radii, while white dotted lines denote DM radii---both expressed in kilometers. In the right panel, the overlaid contours represent the overall stellar radius \( R = \max\ (R_{\rm NM}, R_{\rm DM}) \), also given in kilometers.

The mass contours exhibit non-monotonic trajectories---bending back toward lower central DM densities after reaching a local peak. Along this folded segment of the stability boundary, configurations with the same gravitational mass intersect distinct radius contours, implying the existence of two physically distinct stars with identical mass but markedly different sizes. Additional twin branches also appear near the extremities of the stability domain: one along the vertical segment of the stability boundary (low DM densities), where compact stars with dark cores are supported by extended nuclear mantles, and another along the horizontal segment of the boundary (low nuclear matter densities), where the same-mass configurations are characterized by diffuse DM halos and small nuclear cores. The right panel, in particular, shows that such twin-star pairs can differ in overall radius by several kilometers while remaining gravitationally indistinguishable, reinforcing the multifaceted twin-star phenomenology enabled by two-fluid dynamics. We emphasize that this behavior is not universal across all DM models. In our study, prominent twin branches appear most clearly in specific regimes---such as the case shown here with $m_{\chi} = 1$ GeV and $g_{\chi}/m_{v} = 0.01$ MeV$^{-1}$, and also in another case with $m_{\chi} = 5$ GeV and $g_{\chi}/m_{v} = 0.05$ MeV$^{-1}$ (Fig.~\ref{fig:figure6}). In some intermediate scenarios, such as $m_{\chi} = 5$ GeV and $g_{\chi}/m_{v} = 0.03$ MeV$^{-1}$, a limited number of twin configurations can still be found, but the effect is less pronounced. This suggests that while twin-star solutions are not a generic feature of all two-fluid models, they can emerge under specific conditions tied to the interplay between DM stiffness and coupling strength.

These twin configurations, though degenerate in gravitational mass, are compositionally distinct. The compact, nuclear matter-dominated branch typically features dense nuclear matter extending to radii of 10-12 km, with a small, gravitationally compressed DM core confined to the inner 1-3 km. In contrast, the twin solutions on the DM-dominated branch exhibit dilute nuclear cores embedded within extended dark halos, with DM radii reaching 15-25 km depending on the central density. The transition between these branches does not involve a discontinuity in mass, but a qualitative shift in the spatial distribution of pressure support---from baryon-dominated to DM-dominated structure. See also Fig.~\ref{fig:twin_confirm} in the Appendix~\ref{sec:appendx3} for a visualization of the mass-degenerate branches. This reversal in the fluid hierarchy arises not from microphysical phase transitions, but from the dynamical balance between the two components in general relativistic equilibrium. The result is a genuinely new class of twin-star behavior: one fluid maintains compactness through high-density support, while the other stabilizes a more diffuse configuration via long-range gravitational coupling. Importantly, such configurations would be indistinguishable in mass measurements, yet their radii---and hence tidal deformabilities---could differ substantially, offering a novel observational window into the presence and role of DM in neutron star interiors.

These two-fluid twin-star configurations demonstrate that complex stellar multiplicity can emerge not only from microphysical phase transitions but also from nonlinear fluid coupling in general relativistic equilibrium. Their existence highlights the capacity of DM interactions to reshape the mass-radius landscape of neutron stars in subtle but observable ways. As such, they offer a new paradigm for interpreting future multimessenger data---not as artifacts of phase structure alone, but as signatures of underlying multi-component dynamics within compact stars.
\section{Summary}
\label{sec:5}
In this work, we have conducted a comprehensive analysis of radial oscillations and dynamical stability in two-fluid neutron stars composed of ordinary nuclear matter and a gravitationally coupled DM component. Building on a fully relativistic formalism, we derived and solved the coupled eigenvalue equations for small-amplitude radial perturbations in stars with independently conserved fluids, and systematically charted their stability boundaries with two different methods, i.e, the eigenmode analysis and a generalized turning-point method. By accommodating mirror DM and vector-interacting fermionic DM with self-interactions, we systematically examined how the two-fluid dynamics reshape the equilibrium structure and radial stability of neutron stars.

\begin{figure*}[tbp]
    \centering
    \includegraphics[width=\textwidth]{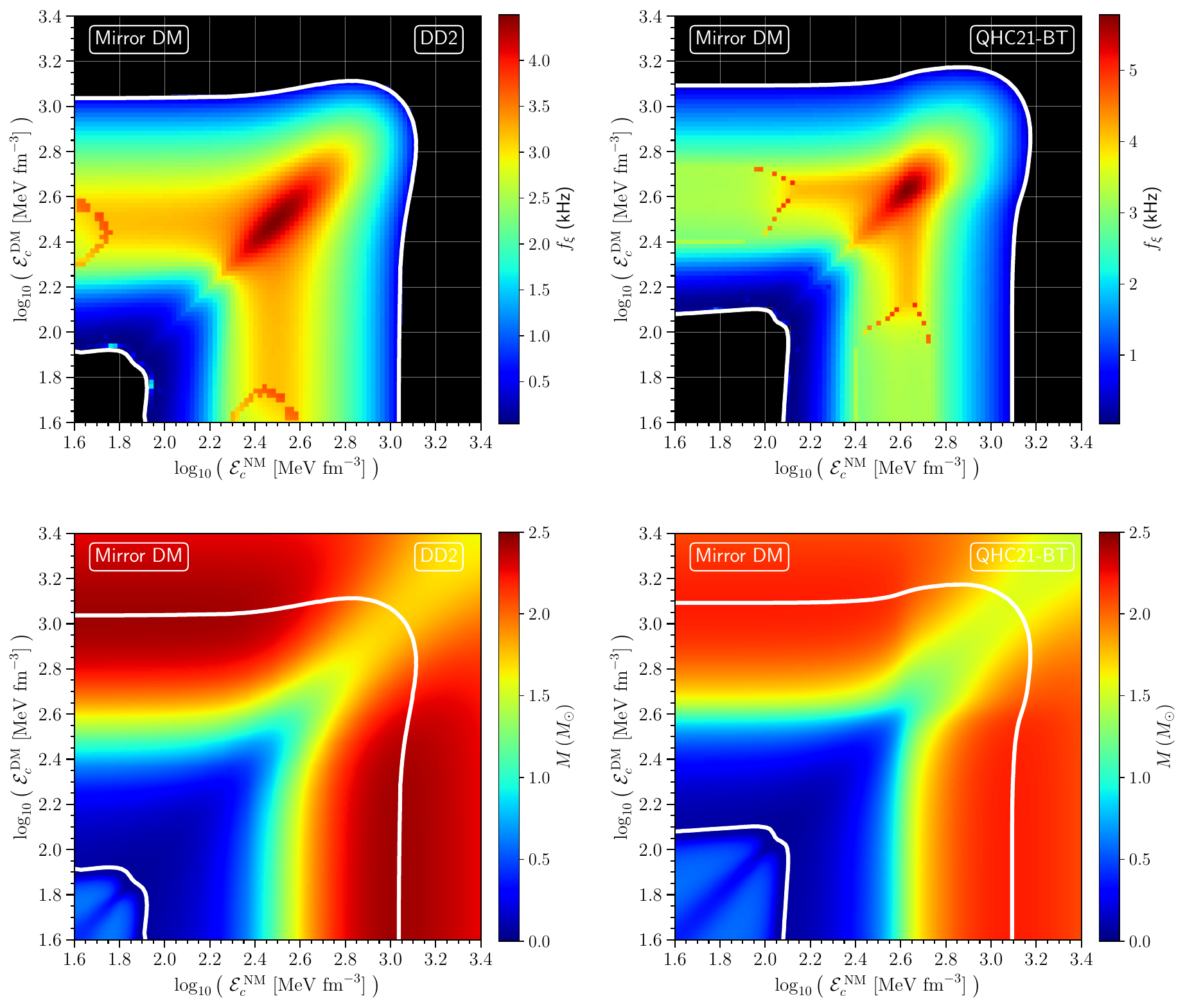} 
    \caption{Two-fluid stability analysis with mirror DM using alternative nuclear EOS. Top row: Fundamental radial mode frequency \( f_{\xi} \) (in kHz) across the \( \log_{10}({\cal E}_{c}^{\rm NM}) \)-\( \log_{10}({\cal E}_{c}^{\rm DM}) \) parameter space, for DD2 (left) and QHC21-BT (right). Bottom row: Total gravitational mass \( M \) (in \( M_{\odot} \)) over the same parameter space. In all panels, the solid white curves trace the dynamical stability boundary as determined independently within each method—via vanishing $f_{\xi}$ (i.e., where $f_{\xi} = 0$) in the top row and via the generalized turning-point criterion in the bottom row. The EOS used for both fluids is identical in each case (mirror matter assumption). These results confirm the presence of two-sided stability boundaries and demonstrate consistent predictions from both approaches across multiple nuclear equations of state.}
    \label{fig:figure10}
\end{figure*}

We have demonstrated that two-fluid stars exhibit a rich spectrum of stable configurations inaccessible in single-fluid systems. These include ultra-dense equilibrium solutions with nuclear central densities exceeding the instability threshold of standard single-fluid neutron stars by factors of two or more, yet maintaining radii in the typical 9-11 km range. Such objects feature stratified interiors with compact dark cores embedded within ultra-dense nuclear mantles, and may serve as natural astrophysical laboratories for probing high-density QCD phenomena. We also identified twin-star-like solutions---pairs of stable configurations with identical gravitational masses but substantially different radii and internal fluid compositions. Unlike conventional twin stars driven by first-order phase transitions, these configurations arise dynamically from nonlinear coupling between fluids in general relativistic equilibrium, and include both dark-core and dark-halo branches depending on the central density balance.

From a methodological standpoint, our study highlights the complementary utility of eigenmode analysis and generalized turning-point diagnostics. While both approaches yield consistent high-density stability boundaries, the turning-point method becomes increasingly sensitive near the low-density (left-edge) boundary, often requiring extremely fine-grained mass and particle number resolution. In such regimes, the eigenfrequency approach offers a more robust and practical diagnostic. However, it remains an open question whether such agreement of tracing similar stability boundaries by both methods persists in more complex scenarios. In single-fluid stars, for example, first-order phase transitions can decouple the location of the mass extremum from the actual onset of radial instability, potentially leading to additional stability branches that are not captured by standard turning-point diagnostics. Exploring whether similar deviations occur in multi-fluid systems---especially in the presence of density discontinuities, additional fluid interactions (entrainment effects), rapid phase transitions or metastable phases---represents an important direction for future work.

Taken together, our findings expand the theoretical landscape of neutron star models by illustrating how two-fluid dynamics can support exotic stable configurations beyond those allowed in conventional scenarios. These results have important implications for multimessenger astrophysics, especially in the interpretation of gravitational wave signals, mass-radius measurements, and post-merger dynamics. Future observational constraints on tidal deformability or quasi-normal mode frequencies may offer key signatures of two-fluid dynamics, providing a new window into the microphysics of dense matter and the possible role of DM in compact star interiors.

\acknowledgments
This work is supported in part by Japan Society for the Promotion of Science (JSPS) KAKENHI Grant Numbers 
JP23K20848  
and JP24KF0090. 
We thank Prof. Toru Kojo for providing the high-density extension of the QHC21-BT equation of state used in this study.
A.K. acknowledges insightful discussions with Prof. Ben Kain during the initial stages of this work, particularly regarding the numerical implementation of two-fluid perturbation equations.
\appendix
\section{Mirror DM with DD2 and QHC21-BT}
\label{sec:appendx1}

To complement the results presented in the main text for the QMC-RMF4 EOS, we perform the stability analysis with two independent methods, using two additional nuclear matter EOSs, i.e., DD2 and QHC21-BT. Fig.~\ref{fig:figure10} summarizes these results under the mirror DM scenario. In the top row, we show the contour maps of the fundamental radial mode frequency $f_{\xi}$, calculated via the eigenmode approach and plotted across the two-dimensional parameter space of central energy densities for nuclear matter and DM. The bottom row shows the corresponding total mass distribution, with white curves marking the dynamical stability boundaries derived using the generalized turning-point method.

Both EOSs exhibit qualitatively similar stability structures to those observed in the QMC-RMF4 case: the stable domain forms a closed island bounded on both the low-density and high-density sides, and the maximum frequency appears along the diagonal, where the two fluids contribute comparably. As before, the agreement between the boundary curves obtained from the two independent methods remains excellent across both DD2 and QHC21-BT, thereby reinforcing the robustness and generality of our two-pronged approach. These results are included here for completeness and provide an additional validation of our framework across different nuclear microphysics models.

In both DD2 and QHC21-BT cases, we also observe a characteristic structure in the frequency maps: elongated ridges of high $f_{\xi}$ values extend outward from the central region toward the horizontal and vertical axes. This behavior mirrors the pattern seen earlier for the QMC-RMF4 EOS (Fig.~\ref{fig:figure1}) and reflects configurations, where one fluid remains dominant while the other is dilute. These high-frequency ridges correspond to effectively single-fluid behavior, where the contribution of the subdominant component is minimal and the dominant fluid alone provides the primary strong restoring force against radial perturbations---highlighting the persistence of this feature across different nuclear EOS microphysics.

\section{Stability boundary curves using QHC21-BT EOS with the self-interacting DM model}
\label{sec:appendx2}
\begin{figure}[htbp]
    \centering
    \includegraphics[width=\columnwidth]{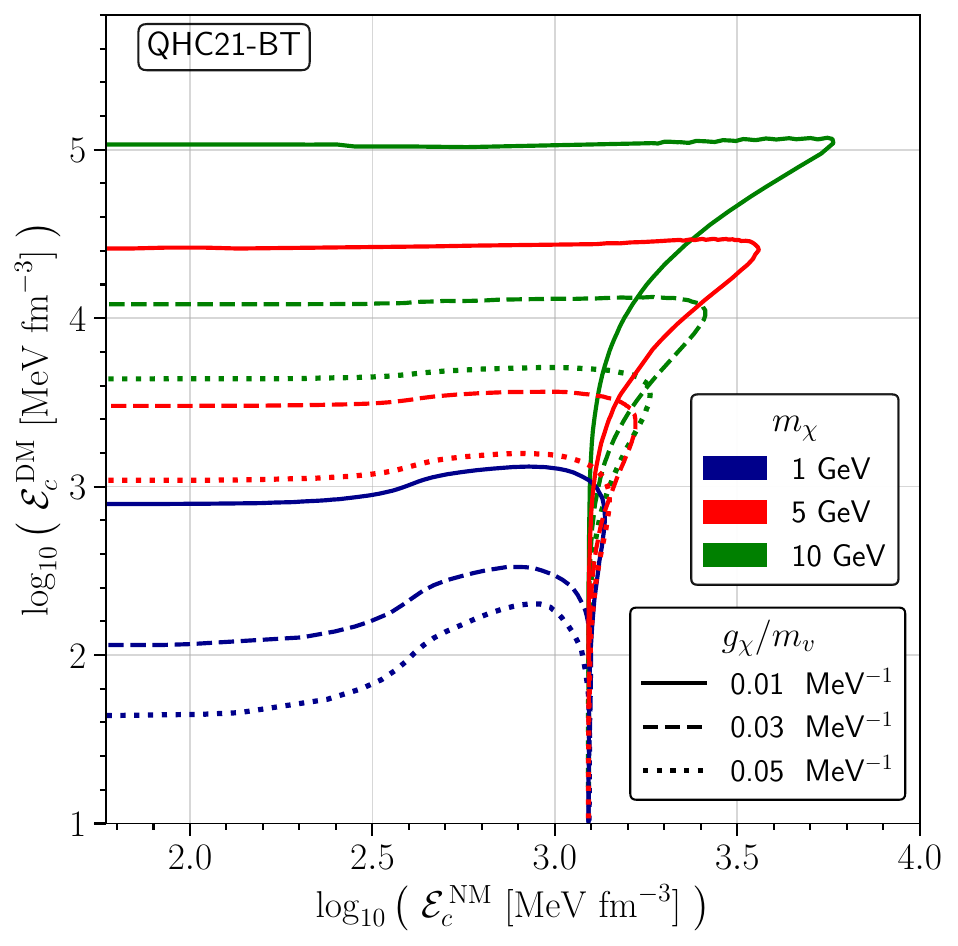}
    \caption{High-density (right-edge) dynamical stability boundaries for two-fluid neutron star configurations using the QHC21-BT as nuclear matter EOS. Each curve corresponds to a specific combination of dark matter mass ($m_{\chi} = 1,\ 5,\ 10$ GeV) and coupling strength ($g_{\chi}/m_{v} = 0.01,\ 0.03,\ 0.05$ MeV$^{-1}$). The stability boundaries are computed using the generalized turning-point method. This figure complements Fig.~\ref{fig:figure7} in the main text, which employed QMC-RMF4 EOS for the nuclear matter component.}
    \label{fig:figure11}
\end{figure}

\begin{figure}[htbp]
    \centering
    \includegraphics[width=\columnwidth]{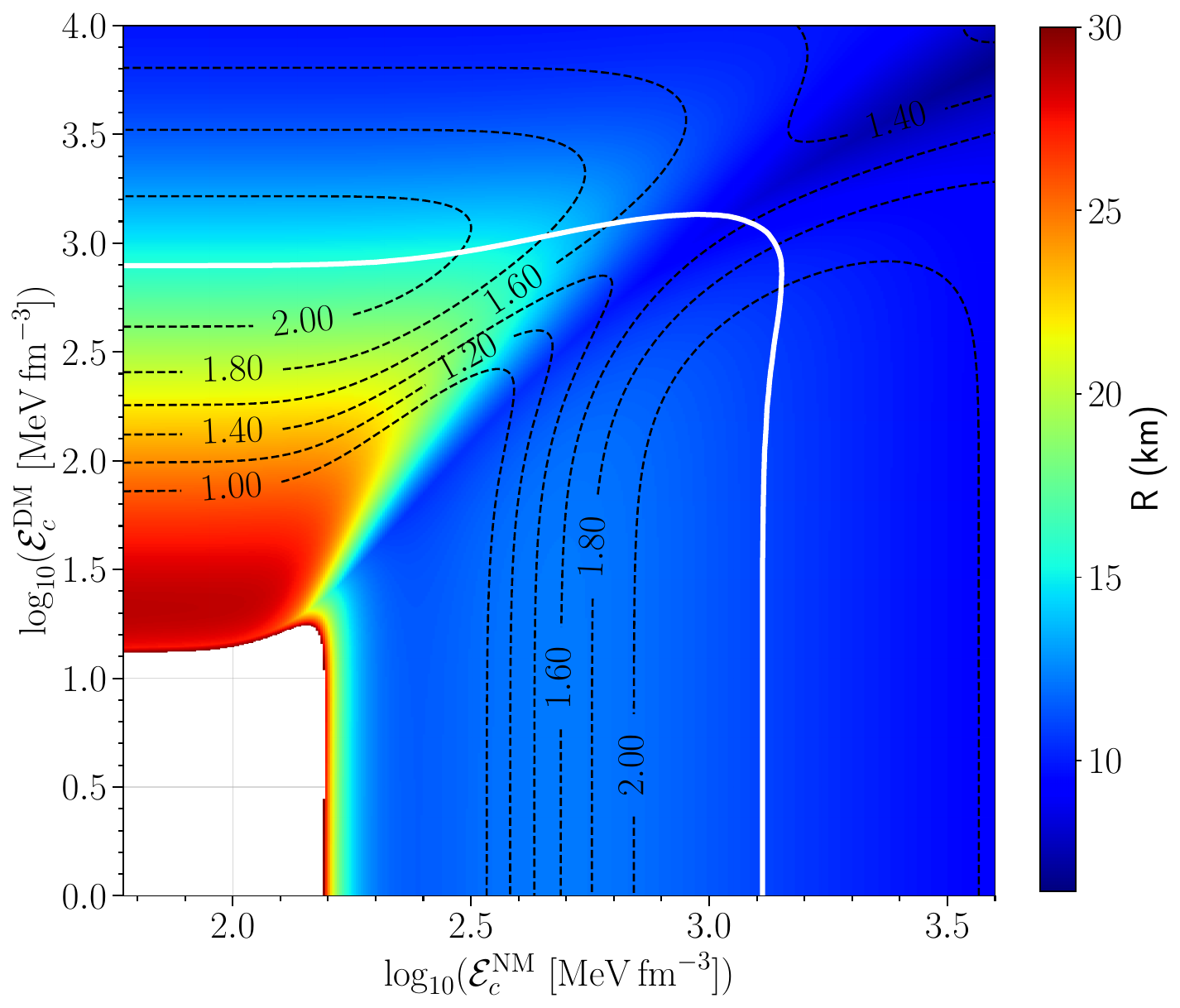}
    \caption{Confirmation of twin-star configurations in the two-fluid model using QMC-RMF4 nuclear matter EOS and vector-interacting dark matter with \(m_{\chi}=1\) GeV and \(g_{\chi}/m_{v}=0.01\) MeV\(^{-1}\). The background color map shows the overall stellar radius (in km), capped at 30 km. Dashed black curves indicate total gravitational mass contours at \(M/M_\odot = 1.00, 1.20, 1.40, 1.60, 1.80, 2.00\). Several mass contours pass through regions with significantly different radii, confirming the existence of physically distinct equilibrium configurations with identical gravitational mass—i.e., twin stars.}
    \label{fig:twin_confirm}
\end{figure}

For completeness, Fig.~\ref{fig:figure11} presents the high-density stability boundaries computed using the QHC21-BT nuclear matter EOS in place of QMC-RMF4. The trends observed here are consistent with those discussed in the main text: increasing dark matter mass softens the EOS, shifting the stability boundary toward higher nuclear densities, while increasing the coupling $g_{\chi}/m_{v}$ stiffens the dark sector and shifts the boundary toward lower DM central densities. Although the absolute positions of the curves differ across nuclear EOS models, the overall morphology and parameter dependence of the stability boundaries remain qualitatively robust.  The inclusion of QHC21-BT---an EOS constructed by smoothly interpolating between hadronic and quark degrees of freedom---helps demonstrate the insensitivity of these trends to the underlying nuclear microphysics.

\section{Explicit confirmation of twin-star behavior}
\label{sec:appendx3}

In Section~\ref{sec:4}, we discussed the emergence of twin-star configurations in the two-fluid framework as an outcome of nonlinear gravitational equilibrium between nuclear matter and dark matter components. While Fig.~\ref{fig:figure9} already illustrates this behavior through intersecting radius contours along fixed-mass tracks, we now include an explicit visualization to reinforce the claim.

In Fig.~\ref{fig:twin_confirm}, we show the total stellar radius (in km) as a background color map over the central density plane \((\log_{10} \mathcal{E}^{\rm NM}_{c}, \log_{10} \mathcal{E}^{\rm DM}_{c})\), overlaid with dashed black contours of constant gravitational mass. The figure corresponds to the same two-fluid model shown in Fig.~\ref{fig:figure9} (QMC-RMF4 nuclear matter EOS and vector-interacting dark matter with \(m_{\chi}=1\) GeV, \(g_{\chi}/m_{v}=0.01\) MeV\(^{-1}\)).

It is clearly evident that multiple constant-mass contours---particularly in the astrophysically relevant range \(M \gtrsim 1\,M_{\odot}\)---span regions with radii differing by several kilometers. This confirms the existence of distinct equilibrium solutions that are gravitationally indistinguishable but compositionally and structurally distinct. We emphasize that this degeneracy is not a numerical artifact, but arises robustly in the two-fluid equilibrium structure, and is consistent with the definition of twin stars used in the literature, even though no first-order phase transition is involved.

This supplemental plot directly supports the interpretation of twin-star behavior presented in the main text.

\bibliographystyle{apsrev4-2}
\bibliography{main.bib} 
\end{document}